\documentclass[a4paper,11pt]{article}

\usepackage{jcappub} 

\usepackage[T1]{fontenc} 
\usepackage{natbib}
\usepackage{amssymb}
\usepackage{amsmath}
\usepackage{amsfonts}
\usepackage{graphicx}
\usepackage[hang,tight,raggedright]{subfigure}
\usepackage{multirow}
\usepackage{subfigure}
\usepackage{float}
\usepackage{graphics}
\usepackage{slashed}
\usepackage{color}
\usepackage{bm}
\usepackage{braket}
\usepackage{psfrag}
\usepackage{booktabs}
\usepackage{latexsym}
\usepackage{pythonhighlight}

\newcommand{\be}{\begin{equation}}
\newcommand{\ee}{\end{equation}}
\newcommand{\besp}{\begin{equation}\begin{split}}
\newcommand{\eesp}{\end{split}\end{equation}}

\title{Gravitational waves from cosmological first-order phase transitions with precise hydrodynamics}

\author[a]{Chi Tian,}
\author[b,1]{Xiao Wang\note{Corresponding author.},}
\author[b]{and Csaba Bal\'azs}

\affiliation[a]{School of Physics and Optoelectronics Engineering, Anhui University, 111 Jiulong Road, Hefei, Anhui, China 230601}
\affiliation[b]{School of Physics and Astronomy, Monash University, Melbourne 3800 Victoria, Australia}

\emailAdd{ctian@ahu.edu.cn}
\emailAdd{xiao.wang1@monash.edu}
\emailAdd{csaba.balazs@monash.edu}

\abstract{
We calculate the gravitational wave spectrum generated by sound waves during a cosmological phase transition, incorporating several advancements beyond the current state-of-the-art.  Rather than relying on the bag model or similar approximations, we derive the equation of state directly from the effective potential.  This approach enables us to accurately determine the hydrodynamic quantities, which serve as initial conditions in a generalised hybrid simulation.  This simulation tracks the fluid evolution after bubble collisions, leading to the generation of gravitational waves.
Our work is the first self-consistent numerical calculation of gravitational waves for the real singlet extension of the standard model.  Our computational method is adaptable to any particle physics model, offering a fast and reliable way to calculate gravitational waves generated by sound waves.  With fewer approximations, our approach provides a robust foundation for precise gravitational wave calculations and allows for the exploration of model-independent features of gravitational waves from phase transitions.
}

\begin{document}
	\maketitle
	\flushbottom

\section{Introduction}

The first detection of gravitational waves (GWs) by the LIGO-Virgo collaboration marked a major breakthrough in astrophysics~\cite{LIGOScientific:2016aoc}.  Since then, detecting the diffuse gravitational wave background has become a key objective.  Recent pulsar timing arrays, such as NANOGrav~\cite{NANOGrav:2023gor}, CPTA~\cite{Xu:2023wog}, EPTA~\cite{EPTA:2023fyk}, and PPTA~\cite{Reardon:2023gzh}, have hinted at the possible discovery of a stochastic gravitational wave background (SGWB) at very low frequencies.  In the near future, space-based observatories like LISA~\cite{LISA:2017pwj,Colpi:2024xhw}, TianQin~\cite{TianQin:2015yph}, Taiji~\cite{Hu:2017mde}, DECIGO~\cite{Kawamura:2011zz}, and BBO~\cite{Corbin:2005ny}, will explore the SGWB across a wide frequency range, from millihertz (mHz) to a few hertz, with unprecedented precision.

To prepare for the influx of new gravitational wave data, precise theoretical models of potential SGWB sources are essential.  While astrophysical processes such as black hole mergers generate GW signals with a characteristic spectrum, early universe processes, including first-order phase transitions (FOPTs), offer distinct GW signatures.  Notably, at temperatures around the TeV scale, corresponding to the electroweak epoch, first-order phase transitions  could generate a peak in the mHz range of the GW spectrum.

However, the standard model (SM) of particle physics predicts only a smooth crossover transition at the electroweak scale~\cite{Kajantie:1996mn,Kajantie:1996qd}.  To accommodate a first-order phase transition, extensions to the standard model, such as the scalar singlet extension~\cite{Espinosa:2007qk,Profumo:2007wc,Espinosa:2011ax,Chiang:2018gsn,Athron:2022jyi}, or the two-Higgs doublet model~\cite{Lee:1973iz,Branco:2011iw,Dorsch:2016nrg,Basler:2017uxn,Fontes:2017zfn,Wang:2019pet}, are required.  GWs produced during such phase transitions offer a unique opportunity to probe these beyond-the-Standard-Model (BSM) theories.

During an FOPT, a stable vacuum is separated from the metastable vacuum by a potential energy barrier.  As the temperature drops, bubbles of the stable low-temperature phase nucleate within the metastable high-temperature phase, driving the FOPT through successive stages of bubble nucleation, expansion, and collision.
Eventually, these bubbles convert the Universe from the unstable phase to a stable phase.
According to recent studies~\cite{Caprini:2015zlo, Caprini:2019egz, Hindmarsh:2020hop, Athron:2023xlk}, there are three primary mechanisms by which gravitational waves can be produced during an FOPT: bubble collisions, sound waves, and turbulence.  For thermal FOPTs without strong supercooling, the gravitational waves generated by sound waves are typically the dominant contribution~\cite{Caprini:2015zlo, Caprini:2019egz, Hindmarsh:2013xza}.

Making precise predictions for phase transition gravitational waves (PTGWs) remains challenging, complicating the task of imposing constraints on BSM theories using future SGWB data.  The conventional framework for predicting PTGWs involves determining phase transition parameters, describing various aspects of the phase transition, and then using these inputs to calculate the PTGWs through fitting formulas.

In recent years, significant efforts have been made to better understand the theoretical uncertainties surrounding these phase transition parameters~\cite{Croon:2020cgk, Athron:2022jyi}, which are derived using different methods.  Key parameters include the energy budget~\cite{Espinosa:2010hh, Leitao:2014pda, Giese:2020rtr, Giese:2020znk, Wang:2020nzm, Wang:2022lyd, Wang:2023jto}, bubble wall velocity $v_w$ \cite{Moore:1995ua, Moore:1995si, Laurent:2020gpg, Laurent:2022jrs, Dorsch:2021nje, Dorsch:2021ubz, Dorsch:2023tss, Wang:2020zlf, Jiang:2022btc, DeCurtis:2022hlx, DeCurtis:2023hil}, reference temperature $T_*$ \cite{Wang:2020jrd,Athron:2022mmm}, and the inverse duration of phase transitions, $\beta$.  The energy budget reflects how much of the energy released during an FOPT is converted into the energy of scalar fields and the kinetic energy of the plasma.  Specifically for sound waves, this energy budget is characterised by the kinetic energy fraction $K$, and is typically quantified by the strength parameter $\alpha$ and the bubble wall velocity $v_w$.

To better model gravitational waves generated by sound waves, several approaches have been developed beyond the conventional \textit{Scalar field} + \textit{fluid lattice simulation} \cite{Hindmarsh:2013xza, Hindmarsh:2015qta, Hindmarsh:2017gnf}.  Semi-analytical methods such as the \textit{Sound shell model} \cite{Hindmarsh:2016lnk, Hindmarsh:2019phv, Guo:2020grp, Wang:2021dwl, Cai:2023guc, RoperPol:2023dzg, Giombi:2024kju} offer more efficient alternatives, along with numerical methods like the \textit{Hybrid simulation} \cite{Jinno:2020eqg, Jinno:2021ury} and the \textit{Higgsless simulation} \cite{Jinno:2022mie, Blasi:2023rqi}.
Crucially, unlike the broken power-law spectrum obtained through fitting formulas, these alternative approaches suggest that PTGWs generated by sound waves should exhibit a double broken power-law feature.

In the absence of a straightforward method to directly connect PTGWs to realistic beyond-the-Standard-Model (BSM) scenarios, we utilise the framework proposed in ref.~\cite{Wang:2024slx} to calculate the gravitational waves generated by sound waves from the Lagrangian of a well-motivated particle physics model.  In this work, we focus on the real scalar singlet extension of the SM with $Z_2$ symmetry, referred to as xSM in the rest of this article.

Our objective is to demonstrate the framework's applicability to more complex and realistic models, emphasising its potential adaptability to any particle physics scenario.  Starting with the effective potential, we first derive the equation of state (EoS), which characterises the hydrodynamics of phase transitions.  Based on this EoS, we perform a detailed analysis of the hydrodynamics specific to the xSM model.  This allows us to precisely determine the feasible hydrodynamic solutions and examine the correlations between the bubble wall velocity and these solutions.

Using these hydrodynamic solutions as initial conditions, we extend the hybrid simulation method to track the fluid evolution after bubble collisions.  The results are then projected onto three-dimensional grids, from which we extract GW spectra.  This approach enables us to compute the SGWB generated by sound waves specific to this particle physics model.

In contrast to previous studies, our approach utilises simplified hydrodynamic simulations based on an EoS directly derived from a particle physics model, rather than relying on specific approximations.  The uncertainties in our method stem only from the effective potential, the nucleation rate, the bubble wall velocity, and the intrinsic uncertainties of the hybrid simulation method.  This approach minimises additional uncertainties associated with the EoS and removes the need for the redundant phase transition parameter $\alpha$. As a result, it offers a more robust framework for calculating PTGWs with fewer approximations, greater accuracy, and easier implementation.  By reducing the reliance on approximations, this framework allows for a more comprehensive analysis of the theoretical uncertainties in PTGWs.  It also opens the door to further exploration of model-independent features of PTGWs generated by sound waves.

This work is organised as follows. In section~\ref{sec:pandeos}, we summarise the conventional perturbative calculation of the effective potential for the $Z_2$-symmetric real singlet extension of the SM, and show the derivation of the equation of state for this model.
Section~\ref{sec:Bnucl} gives a brief review of our construction of the bubble nucleation history.
In section~\ref{sec:Hydro}, with the EoS of this specific particle physics model, we analyse the relativistic hydrodynamics of the FOPT, presenting different hydrodynamic solutions before and after bubble collisions. We also illustrate the relationship between the bubble wall velocity and the hydrodynamic solutions.
In section~\ref{sec:GWs}, we show the technical details used to calculate the GW spectra, and then present the GW spectra for benchmark points of the real singlet extension of the SM.
Final discussion and conclusions are given in section~\ref{sec:concl}.

\section{Effective potential and the equation of state}
\label{sec:pandeos}

Cosmological first-order phase transitions are typically considered in the context of a well-motivated particle physics model.
In this study, we use the real singlet extension of the SM with $Z_2$ symmetry to demonstrate the use of our methodology.
The tree-level scalar potential of this model is
\begin{equation}
    V(H, S) = -\mu_h^2H^\dagger H + \lambda_h(H^\dagger H)^2 - \frac{\mu_s^2}{2} S^2 + \frac{\lambda_s}{4}S^4 + \frac{\lambda_{hs}}{2}H^\dagger HS^2\,,
    \label{eq:treeHp}
\end{equation}
where the Higgs doublet and the $Z_2$-symmetric real singlet are
\begin{equation}
    H \equiv \frac{1}{\sqrt{2}} \left(\begin{array}{c} G^\pm\\
\varphi_h + h + iG^0
\end{array}\right)\, \quad \mathrm{and} \quad S \equiv \varphi_s + s\, .
\end{equation}
Here $\varphi_h$ and $\varphi_s$ are background fields, $h$ is the Higgs boson, and $G^{\pm,0}$ are Goldstone bosons.
In general, both $h$ and $s$ can obtain non-zero vacuum expectation value (VEV) at zero temperature.
Here, for simplicity, we assume that only $h$ can develop a non-zero VEV at zero temperature, and we have 
\begin{equation}
    \varphi_h|_{T=0} = v_0 \approx 246.22~\mathrm{GeV}, \quad \varphi_s|_{T=0} = 0\,\,.
\end{equation}
According to this, the masses of these scalar bosons can be derived as
\begin{align}
m_h^2 &= -\mu_h^2 + 3\lambda_hv_0^2 = 2\lambda_h v_0^2,\notag\\
m_s^2 &= -\mu_s^2 + \frac{1}{2}\lambda_{hs}v_0^2,\notag\\
m_{G^\pm}^2 &= -\mu_h^2 + \lambda_{h}v_0^2 = 0,
\end{align}
and we have
\begin{align}
\mu_h^2 &= \frac{m_h^2}{2},\quad\mu_s^2 = -m_s^2 + \frac{1}{2}\lambda_{hs}v_0^2.
\end{align}
In this model, there are thus three free parameters that are denoted by $m_s$, $\lambda_s$, and $\lambda_{hs}$.
With this specific model, we are able to write down the effective potential and construct the EoS for the electroweak FOPT in the following subsections.

\subsection{Effective potential}

In this work, we employ the conventional 4D perturbative method~\cite{Quiros:1999jp} to calculate the effective potential.
To obtain a more robust result, one could use the 3D dimensional reduction effective field theory~\cite{Farakos:1994kx,Kajantie:1995dw} or lattice simulations~\cite{Kajantie:1993ag,Farakos:1994xh,Kajantie:1995kf} to derive the corresponding effective potential.
With the 4D perturbative method, the one-loop effective potential can be expressed as
\begin{equation}
V_{\rm eff}(\varphi,T) \equiv V_{\rm tree}(\varphi) + V_{\rm CW}(\varphi) + V_{\rm T}(\varphi,T),
\label{eq:Veff}
\end{equation}
where $\varphi \equiv (\varphi_h, \varphi_s)$, $V_{\rm tree}(\varphi)$ is the tree level effective potential, $V_{\rm CW}(\varphi)$ is the Coleman-Weinberg potential, and $V_{\rm T}(\varphi,T)$ is the one-loop thermal correction.
Based on the tree-level scalar potential eq.~\eqref{eq:treeHp}, the tree level effective potential can be written as
\begin{equation}
V_{\rm tree}(\varphi_h,\varphi_s) = -\frac{1}{2}\mu_h^2\varphi_h^2 - \frac{1}{2}\mu_s^2\varphi_s^2 + \frac{\lambda_h}{4}\varphi_h^4 + \frac{\lambda_s}{4}\varphi_s^4 + \frac{1}{4}\lambda_{sh}\varphi_h^2\varphi_s^2.
\end{equation}
To enforce the one-loop corrected masses and mixing angles to be equal to the tree-level values, we use the on-shell renormalization prescription.
Hence the corresponding Coleman-Weinberg potential is
\begin{equation}
    V_{\rm CW} = \sum_i \frac{n_i}{64\pi^2} \left\{  m_i^4(\varphi) \left(\ln\frac{m_i^2(\varphi)}{m_i^2(v_0)} - \frac{3}{2}\right) + 2m_i^2(\varphi)m_i^2(v_0) \right\}\,,
\end{equation}
where $m_i(v_0)$ represents the mass of particle species $i$ at zero-temperature, $n_i$ is the number of degree of freedom (dof) for each particle species, and $m_i(\varphi)$ denotes the finite-temperature field-dependent mass of different particles given in Appendix~\ref{App:mass}.
Note that the Goldstone bosons could cause the IR divergence in the on-shell scheme.
However, ref.~\cite{Chiang:2018gsn} showed that the contribution of Goldstone bosons is small.
Therefore, for simplicity, we neglect the contribution of the Goldstone bosons.
Alternatively, one can adopt the prescription given in Ref.~\cite{Chiang:2018gsn} to circumvent this issue. 

For the one-loop thermal corrections,  the conventional 4D method yields
\begin{equation}
    V_{\rm T} (\varphi, T) = \sum_F \frac{T^4}{2\pi^2}n_F J_F\left(\frac{m_F^2(\varphi)}{T^2}\right) + \sum_B \frac{T^4}{2\pi^2}n_B J_B\left(\frac{m_B^2(\varphi)}{T^2}\right)\,,
\end{equation}
where the subscript $B$ and $F$ denote the bosons and fermions respectively, $n_{F/B}$ represent the d.o.f of bosons and fermions, and the thermal functions are
\begin{equation}
    J_{B/F} = \int_0^\infty dxx^2 \log\left[1\mp e^{- \sqrt{x^2 + m^2/T^2}}\right],
\end{equation}
with $-$ for bosons and $+$ for fermions.
Here, we use the Parwani scheme~\cite{Parwani:1991gq} to incorporate the contribution of daisy resummation by the following replacement
\begin{equation}
    m_B^2 \to \overline{m}_B^2 = m_B^2 + \Pi_B\,,
\end{equation}
where $\Pi_B$ is the thermal mass of bosons, which can be found in Appendix~\ref{App:mass}.

\subsection{The equation of state}

According to thermal field theory, we have the following relation~\cite{Laine:2016hma}
\begin{equation}
    \mathcal{F}(T) = V_{\rm eff}(\varphi_m(T), T) + \mathcal{O}\left(\frac{\ln\mathcal{V}}{\mathcal{V}}\right)\,,
\end{equation}
where $\mathcal{F}(T)$ is the free energy density of the thermodynamic system, $\mathcal{V}$ is the volume of the system, and $\varphi_m(T)$ is the minimum of the effective potential $V_{\rm eff}$ at a specific temperature $T$.
Therefore, in the thermodynamic limit $\mathcal{V}\to \infty$, the computation of $\mathcal{F}(T)$ reduces to calculating the effective potential $V_{\rm eff}$ and tracing its minimum.
Using the free energy density, the pressure $p$, the energy density $e$, the enthalpy $w$ and the entropy density $s$ can be defined as
\begin{equation}
    p = -\mathcal{F}, \quad e = T\frac{\partial p}{\partial T} - p, \quad w = T\frac{\partial p}{\partial T}, \quad s = \frac{\partial p}{\partial T}\, ,
\end{equation}
where we have the relation $w = p+e$.
With these relations, the EoS of the xSM can be obtained as 
\begin{equation}
    p(T) = - V_{\rm eff}\left(\varphi_m(T), T\right), \quad e(T) = -T\frac{d V_{\rm eff}}{d T} + V_{\rm eff}\left(\varphi_m(T), T\right).
\end{equation}

It is well known that in the xSM, one- or two-step phase transitions are possible.
For one-step phase transitions, the EoS of both phases are
\begin{equation}
\begin{split}
    p_+(T) = \frac{1}{3}a_+T^4, &\quad e_+(T) = a_+T^4,\\
    p_-(T) = - V_{\rm eff}\left(\varphi_m^-(T), T\right), &\quad e_-(T) = -T\frac{d V_{\rm eff}}{d T} + V_{\rm eff}\left(\varphi_m^-(T), T\right),
\end{split}\label{eq:EoS1s}
\end{equation}
where the subscripts $+$ and $-$ denote the high-temperature phase (or the symmetric phase) and the low-temperature phase (or the broken phase) respectively, and $a = g_*\pi^2/30$.
Here $g_*$ is the total number of degrees of freedom, which we set $g_*\approx106.75$ for the electroweak phase transition.
In this work, we are interested in two-step phase transitions in the xSM.
For these two-step phase transitions, the first step is a second-order phase transition, while the second step is a first-order phase transition.
Therefore, the EoS for the second step phase transition should be
\begin{equation}
\begin{split}
    p_+(T) = - V_{\rm eff}\left(\varphi_m^+(T), T\right), \quad e_+(T) = -T\frac{d V_{\rm eff}}{d T} + V_{\rm eff}\left(\varphi_m^+(T), T\right)\\
    p_-(T) = - V_{\rm eff}\left(\varphi_m^-(T), T\right), \quad e_-(T) = -T\frac{d V_{\rm eff}}{d T} + V_{\rm eff}\left(\varphi_m^-(T), T\right)
\end{split}\label{eq:EoSf}
\end{equation}
where $\varphi_m^+(T)$ is the local minimum of the effective potential which indicates the metastable high-temperature phase, and $\varphi_m^-(T)$ is the global minimum of effective potential that represents the stable low-temperature phase.

Based on the numerical simulation of the scalar-fluid coupled system~\cite{Kurki-Suonio:1995yaf}, one can assume that $\varphi_m(T)$ does not change with temperature. 
For this approximation, $\varphi_m(T)$ is equal to the constant $\varphi_m^\pm$, which is the local/global minimum of the effective potential at temperature $T_\pm$.
Here $T_+$ is the temperature just in front of the bubble wall, $T_-$ is the temperature just behind the bubble wall.
We can construct another EoS as the following:
\begin{equation}
    p_\pm(T) = - V_{\rm eff}\left(\varphi_m^\pm, T\right), \quad e_\pm(T) = -T\frac{d V_{\rm eff}}{d T} + V_{\rm eff}\left(\varphi_m^\pm, T\right).
\label{eq:EoSa}
\end{equation}
The temperature of the fluid shell surrounding the bubble wall, however, should change within the shell as we shall find in the hydrodynamic analysis shown in section~\ref{sec:Hydro}. 
The above approximation for the EoS might introduce some uncertainties to the hydrodynamic calculation. We will explore the effect of this approximation in our future study.

\section{Bubble nucleation}
\label{sec:Bnucl}

In addition to the effective potential and the EoS, which govern the hydrodynamics during the FOPTs, understanding the nucleation history of bubbles is also crucial for quantifying the resultant GW spectra. 
The bubble nucleation rate can be described by
\begin{equation}
    \Gamma (t) = \Gamma_0 e^{-S_E}\,,
\end{equation}
where $S_E$ is the bounce action. 
For a thermal phase transition, one can use the following approximation $S_E \approx S_3/T$, where $S_3$ is $O(3)$-symmetric bounce action, which is 
\begin{equation}
    S_3 = 4\pi\int_0^\infty drr^2\left[\frac{1}{2}\left(\frac{d\varphi}{dr}\right)^2  + V_{\rm eff}\right].
\end{equation}
This can be derived by the following equation of motion of the background field and boundary conditions:
\begin{equation}
    \frac{d^2\varphi}{d r^2} + \frac{2}{r}\frac{d\varphi}{dr} = \frac{\partial V_{\rm eff}}{\partial \varphi}, \quad \varphi'(0) = 0, \quad \varphi(\infty) = \varphi_{\rm false},
\end{equation}
where $\varphi_{\rm false}$ is the VEV of the high-temperature phase.
In this work, we need to solve the bounce equations for two background fields.
There are various public numerical codes, such as $\mathtt{BubbleProfiler}$~\cite{Athron:2019nbd}, $\mathtt{FindBounce}$~\cite{Guada:2020xnz}, etc., to solve the bounce equation. In this work, we employ the $\mathtt{cosmoTransitions}$~\cite{Wainwright:2011kj} package for this purpose.
Recent advances in the precise calculation of bubble nucleation rates~\cite{Gould:2021ccf,Lofgren:2021ogg,Hirvonen:2021zej,Ekstedt:2021kyx,Ekstedt:2022tqk,Ekstedt:2022ceo,Chala:2024xll},
enable the incorporation of higher-order corrections, thereby reducing theoretical uncertainties.
Furthermore, a new public package, $\mathtt{BubbleDet}$~\cite{Ekstedt:2023sqc}, facilitates systematic treatment of the prefactor $\Gamma_0$.
These improvements will be addressed in our future work.

After nucleation, bubbles expand and collide.
Consequently, the fraction of the high-temperature phase decreases as the volume of the bubbles increases. This can be described by the decreasing fraction of the unstable high-temperature phase $h$, which is defined by
\begin{equation}
h \equiv V_{\rm s}/V_{\rm tot},
\end{equation}
where the total volume $V_{\rm tot}$ is the sum of the volume of the unstable high-temperature phase $V_{\rm s}$, and the volume of the stable low-temperature phase  $V_{\rm b}$, 
The reduction of the volume of the high-temperature phase between times $t$ and $t+dt$ due to the growth of bubbles nucleated between $t'$ and $t' + dt'$ can thus be written as
\begin{equation}
d^2V_{\rm s}(t, t') = -dN_b(t')4\pi R_b^2dR_b\frac{V_{\rm s}(t)}{V_{\rm s}(t')},
\end{equation}
where $dN_b$ represents the number of bubbles nucleated between $t'$ and $t' + dt'$, and $R_b$ denotes the radius of those bubbles at time $t$.
We can write the radius and its derivative
\begin{equation}
    R_b = v_w(t -t'), \quad dR_b = v_w dt,
\end{equation}
in terms of a constant wall velocity, $v_w$, since the bubble wall would reach the steady state very soon after nucleation.
The wall velocity can be determined by the friction between the bubble wall and the surrounding plasma.
Note that the factor $V_s(t)/V_s(t')$ is less than $1$,  indicating that only parts of the bubbles expand into the unstable high-temperature phase, which will change the volume of the high-temperature phase.
Further, the number of bubbles nucleated in the time interval $[t', t' + dt']$ is
\begin{equation}
dN_b = \Gamma(t')V_{\rm s}(t')dt'.
\label{eq:Nb}
\end{equation}
The change of the volume of the high-temperature phase is then
\begin{equation}
dV_{\rm s}(t) = -v_wV_{\rm s}(t)dt\int_{t_c}^{t}dt'\Gamma(t')4\pi v_w^2(t - t')^2.
\end{equation}
The lower limit of the integral reflects that the nucleation rate $\Gamma$ is non-vanishing only below the critical temperature $T_c$.
In other words, $t_c$ is the time at which the temperature is equal to $T_c$.
Dividing by $V_{\rm tot}$, we can derive the differential equation for the fraction of high-temperature phase as
\begin{equation}
\frac{dh}{dt} = -v_wh(t)\int_{t_c}^{t}dt'\Gamma(t')4\pi v_w^2(t - t')^2 .
\end{equation}
Solving the above equation gives 
\begin{equation}
h(t) = \exp\left[-\frac{4\pi}{3}v_w^3\int_{t_c}^{t}dt'\Gamma(t')(t - t')^3\right].
\end{equation}

Given that the typical duration of the FOPT is short, $h(t)$ is rapidly changing and the Euclidean action $S_E$ is dominated by its value around $t=t_*$.
Indeed, by Taylor expanding the bounce action $S_E$ we obtain
\begin{equation}
    S_E(t) \approx S_* - \beta(t - t_*) + \mathcal{O}((t - t_*)^2)  + \ldots,
\end{equation}
where $S_* = S_E(t_*)$, and $\beta$ is the inverse duration of phase transition defined by
\begin{equation}
    \beta = -S_E'(t_*) = -\frac{d S_E}{dt}\bigg|_{t=t_*} = HT\frac{dS_E}{dT} \bigg|_{T=T_*}\, .
\end{equation}
Here we used the adiabatic time-temperature relation $dt = -dT/(TH)$, and $H$ is the Hubble rate.
Then the bubble nucleation rate can be approximated as
\begin{equation}
    \Gamma (t) \approx \Gamma_* \exp[\beta(t-t_*)]\,,
    \label{eq:gam1}
\end{equation}
where $\Gamma_* = \Gamma_0 e^{-S_E(t_*)}$. 
With this, the fraction of the high-temperature phase becomes
\begin{equation}
    h(t) = \exp\left[8\pi\frac{v_w^3}{\beta^4}\Gamma_0e^{-S_* + \beta(t - t_*)}\right]\,.
\end{equation}
Defining $t_f$ such that $h(t_f) = 1/e$, we have 
\begin{equation}
    h(t) = \exp\left[-e^{\beta(t - t_f)}\right]\,,
\end{equation}
where $t_f$ is determined by the following equation
\begin{equation}
    8\pi\frac{v_w^3}{\beta^4}\Gamma_0e^{-S_* + \beta(t_f - t_*)} = 1\,.
    \label{eq:vbeta}
\end{equation}
Because bubbles can only nucleate in the high-temperature phase, dividing eq.~\eqref{eq:Nb} with $V_{\rm tot}$, we can obtain an equation governing the bubble number density:
\begin{equation}
\frac{dn_b}{dt} = \Gamma(t)h(t).
\end{equation}
This yields the asymptotic bubble number density:
\begin{equation}
    n_b = \frac{\Gamma_0}{\beta}e^{-S_* + \beta(t_f - t_*)} = \left(8\pi\frac{v_w^3}{\beta^3}\right)^{-1}\,.
    \label{eq:nb}
\end{equation}
Defining the mean bubble separation as $R_* \equiv n_b^{-1/3}$,
from eq.~\eqref{eq:nb}, we have the following relation 
\begin{equation}
    R_* = (8\pi)^{1/3}\frac{v_w}{\beta}.
    \label{eq:mbR}
\end{equation}
In section~\ref{sec:GWs}, we show that the mean bubble separation is an essential ingredient of the gravitational wave spectrum calculation.
In this work, we choose $t_* = t_f$.
Then with eqs.~\eqref{eq:gam1} and \eqref{eq:vbeta}, the nucleation rate can be approximated as 
\begin{equation}
    \Gamma (t) \approx \beta^4\exp{\left[\beta(t - t_*)\right]}\,.
    \label{eq:AppNR}
\end{equation}
We use this approximation in following calculation, for simplicity.

In practice, we choose $t_*$ as the time when the temperature is equal to the nucleation temperature $T_n$, and we use $\Gamma/H^4 = 1$ to derive the nucleation temperature $T_n$.
As we shall see, the nucleation temperature will serve as an important boundary conditions for the hydrodynamic analysis in the next section.
For more precise results, one can adopt the percolation temperature as the reference temperature, as suggested by Refs.~\cite{Athron:2022mmm, Wang:2020jrd}.

\section{Hydrodynamics} 
\label{sec:Hydro}

Recent studies indicate that sound waves provide the dominant contribution to gravitational waves generated during phase transitions. 
These sound waves result from the collision of fluid shells surrounding bubble walls, with their dynamics determined by the hydrodynamics of the phase transition. 
To obtain a robust prediction for gravitational waves, it is therefore crucial to keep track of the hydrodynamics for a given model.

Following refs.~\cite{Gyulassy:1983rq,Enqvist:1991xw,Laine:1993ey,Kurki-Suonio:1995rrv,Megevand:2012rt,Espinosa:2010hh},
we perform a detailed analysis to determine the allowed hydrodynamic modes.
Based on the general consideration of energy-momentum conservation and increasing entropy, we apply this analysis for the EoS during the electroweak phase transition.
We assume that the bubble wall is planar, which is a reasonable approximation at the steady state.
Then in the rest frame of the expanding bubble wall, the energy-momentum conservation across the bubble wall gives the equations:
\begin{equation}
    v_+v_- = \frac{p_- - p_+}{e_- - e_+}, \quad \frac{v_-}{v_+} = \frac{e_+ + p_-}{e_- + p_+},
    \label{eq:mcond}
\end{equation}
where the subscripts $\pm$ refer to quantities of the high-temperature phase and quantities of the low-temperature, respectively.
To derive the physically possible values of $T_+$ and $T_-$, we enforce $0\leq v_+^2\leq 1$, $0\leq v_-^2 \leq 1$, and the increasing entropy condition gives
\begin{equation}
    \Delta s_{\perp} = s_-\gamma_-v_- - s_+\gamma_+v_+ \ge 0\,,
\end{equation}
which can be rewritten as
\begin{equation}
    \frac{s_-}{s_+} \ge \frac{T_-}{T_+}\frac{e_- + p_+}{e_+ + p_-}.
\end{equation}
To analyse the possible solutions on the $T_+$-$T_-$ plane, it is convenient to use the lines
\begin{equation}
    v_+ = v_- = 0, \quad v_+ = v_- = 1, \quad \Delta s_\perp = 0 .
    \label{eq:vline}
\end{equation}
We also need the Jouguet line:
\begin{equation}
    v_- = c_{s,-},
    \label{eq:jline}
\end{equation}
which indicates that the speed of the flow outside of the bubble wall is equal to the sound speed of the low-temperature phase.
Obtaining solutions can only be done numerically for the EoS in~\eqref{eq:EoSf}.

To obtain an intuitive understanding about the possible solutions in the $T_+$-$T_-$ plane, we introduce the bag model of EoS
\begin{equation}
\begin{split}
    p_+ = \frac{1}{3}a_+T_+^4 - \epsilon\,\,, &\quad e_+ = a_+T_+^4 + \epsilon\,\,,\\
    p_- = \frac{1}{3}a_-T_-^2\,\,, &\quad e_- = a_-T_-^4\,\,.
\end{split}
\end{equation}
This simplified model has analytical solutions that can be written as
\begin{equation}
\begin{split}
    &v_+ = v_- = 0: \quad y^2 = r x^2 + 3\alpha_c\,\,,\\
    &v_+ = v_- = 1: \quad y^2 = r x^2 - 3\alpha_c\,\,,\\
    &\mathrm{Entropy~increse}: \quad r x(y^2 + r x^2/3 + \alpha_c) - y(r x^2 + y^2/3 - \alpha_c) \ge 0\,\,,\\
    &\mathrm{Jouguet}: \quad y^2 = \alpha_c + rx^2 \pm 2\sqrt{\alpha_c^2 + 2r \alpha_c x^2}\,\, .
    \label{eq:bagTT}
\end{split}
\end{equation}
Here we have defined
\begin{equation}
    x \equiv \frac{T_-^2}{T_c^2}, \quad y \equiv \frac{T_+^2}{T_c^2}, \quad r \equiv \frac{a_-}{a_+}, \quad \alpha_c \equiv \frac{\epsilon}{a_+T_c^4} ,
\end{equation}
where $T_c$ is the critical temperature. 

\begin{figure}[t!]
    \centering
    \includegraphics[width=0.5\textwidth]{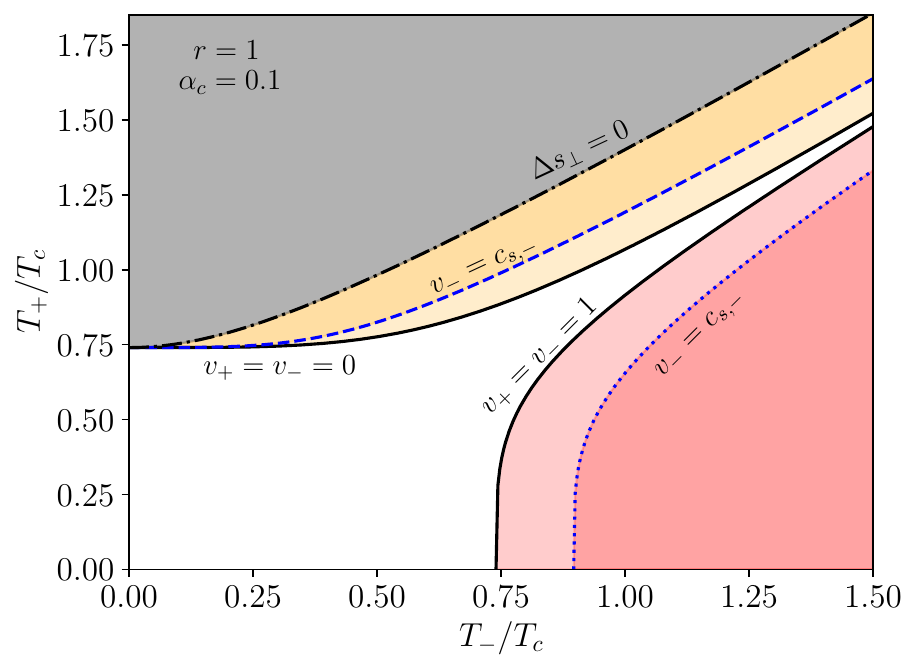}%
    \includegraphics[width=0.5\textwidth]{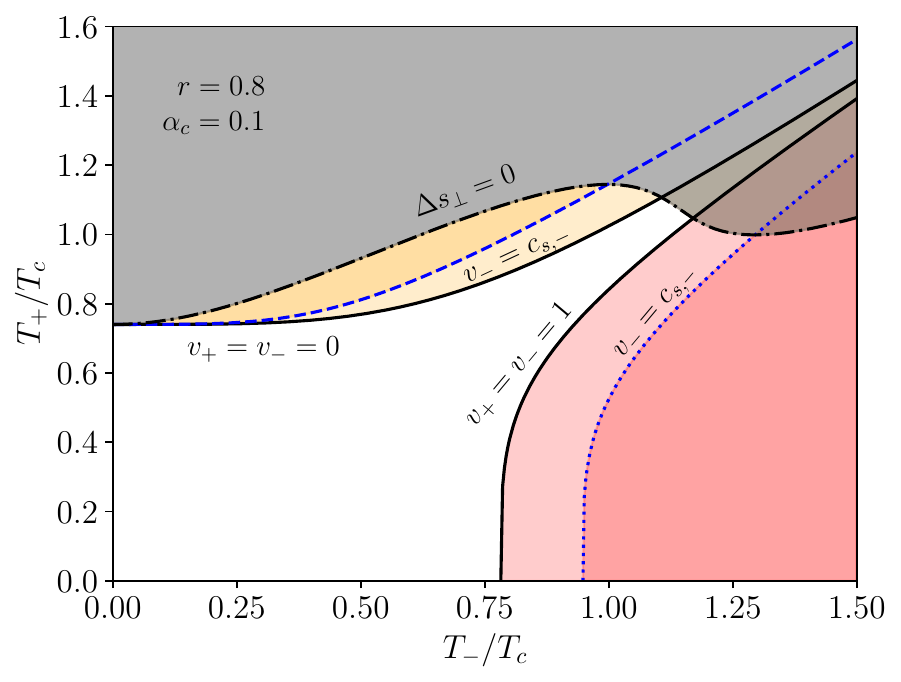}
    \caption{Plots of the curves defined by eq.~\eqref{eq:bagTT} for $\alpha_c=0.1$ in the $T_+$-$T_-$ plane of the bag EoS. The left panel corresponds to $r=1$, and the right panel is for $r=0.8$. 
    The grey shaded regions are excluded by entropy production. 
    The region lying below the $\Delta s_\perp = 0$ line (black dashed) and above the $v_+=v_-=0$ line (black solid) represents deflagrations.
    The dark orange sub-region marks strong deflagrations, and the light orange sub-region marks weak deflagrations.
    The region lying below the $\Delta s_\perp = 0$ line and to the right of the $v_+=v_-=1$ line (black solid) covers detonations.
    The dark red sub-region marks strong detonations, and the light red sub-region marks weak detonations.
    The blue dashed and blue dotted lines are the Jouguet lines.}
    \label{fig:bag}
\end{figure}

The curves given by~\eqref{eq:bagTT} are shown in Fig.~\ref{fig:bag} for $\alpha_c = 0.1$.
The left panel displays results for $r=1$ and the right panel for $r = 0.8$.
Lowering $r$ can alter the $\Delta s_\perp = 0$ line.
From Fig.~\ref{fig:bag}, we can immediately read that there are two branches of solutions for eqs.~\eqref{eq:mcond},
\begin{equation}
    \mathrm{Deflagrations}:~v_- > v_+,\quad \mathrm{Detonaions}:~v_+ > v_-\,\,.
\end{equation}
Here, we choose a convention in which both $v_-$ and $v_+$ are positive.
In both panels of Fig.~\ref{fig:bag}, deflagrations lie below the $\Delta s_\perp = 0$ line (black dashed) and above the $v_+=v_-=0$ line (black solid), whereas detonations
lie below the $\Delta s_\perp = 0$ line and to the right of the $v_+=v_-=1$ line (black solid).
The region of deflagrations is divided into two sub-regions by the blue dashed Jouguet line, and these two sub-regions are depicted by light orange and dark orange, respectively.
The region of detonations is also divided into two sub-regions by the blue dotted Jouguet line, and these sub-regions are represented by light red and dark red.
We shall discuss the difference between these sub-regions in detail in the next subsection.

\begin{table}[!t]
	\centering
	\setlength{\tabcolsep}{2.5pt}
	\renewcommand{\arraystretch}{1.2}
	\begin{tabular}{|c|c|c|c|c|c|c|c|c|c|}
		\hline
		&$m_s\rm~[GeV]$ & $\lambda_s$ & $\lambda_{hs}$ & $T_n~\mathrm{[GeV]}$ & $T_{\rm min}~\mathrm{[GeV]}$ & $T_{\rm max}~\mathrm{[GeV]}$ & $\beta~\mathrm{[s^{-1}]}$ & $\beta/H_*$ & $\alpha_n$\\
		\hline
        $\mathrm{BP}_1$ & $80$ & $1.2$  & $0.82$ & $84.344$ & $79.085$ & $113.632$ & $3.581\times10^{13}$ & $2358.224$ & $0.022$\\
        \hline
        $\mathrm{BP}_2$ & $80$ & $1.2$ & $0.89$  & $53.371$ & $49.01$ & $111.539$ & $5.794\times10^{12}$ & $953.267$ & $0.114$\\
        \hline
	\end{tabular}
	\caption{Benchmark points of the $Z_2$-symmetric xSM and corresponding phase transition parameters.}
	\label{tb:bps}
\end{table}

\begin{figure}[!ht]
    \centering
    \includegraphics[width=0.5\textwidth]{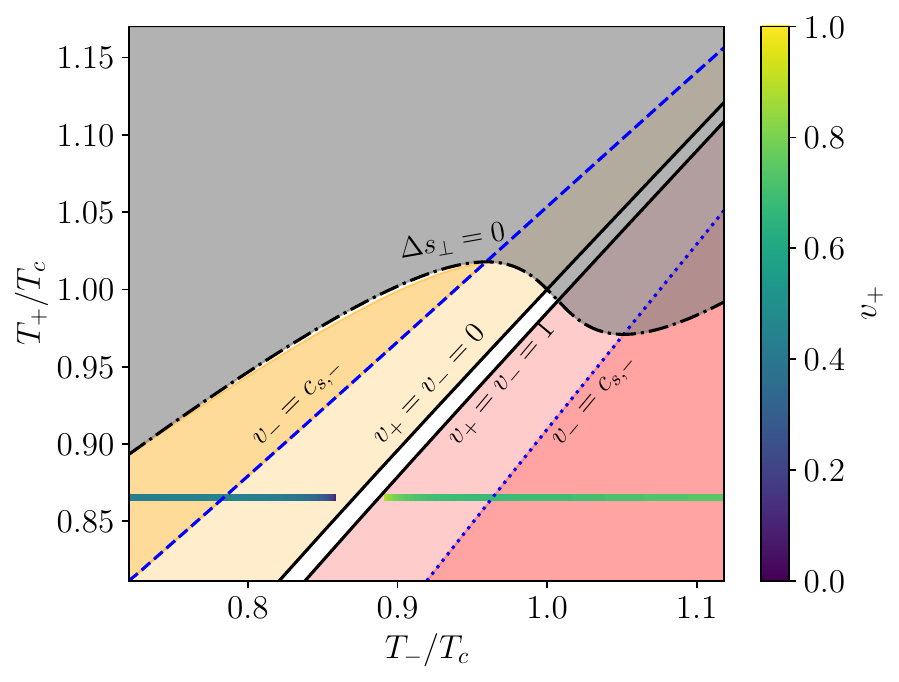}%
    \includegraphics[width=0.5\textwidth]{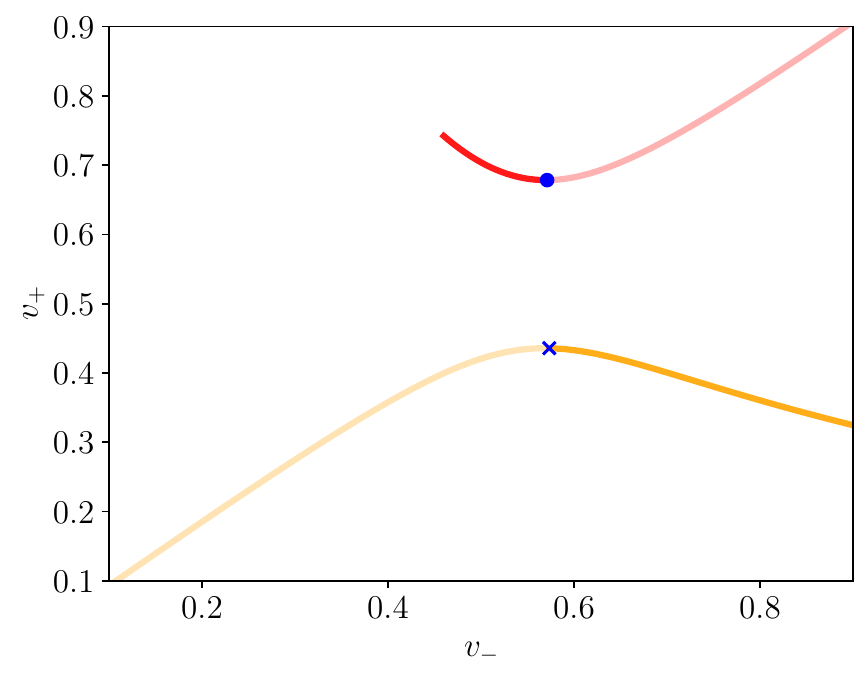}
    
    \includegraphics[width=0.5\textwidth]{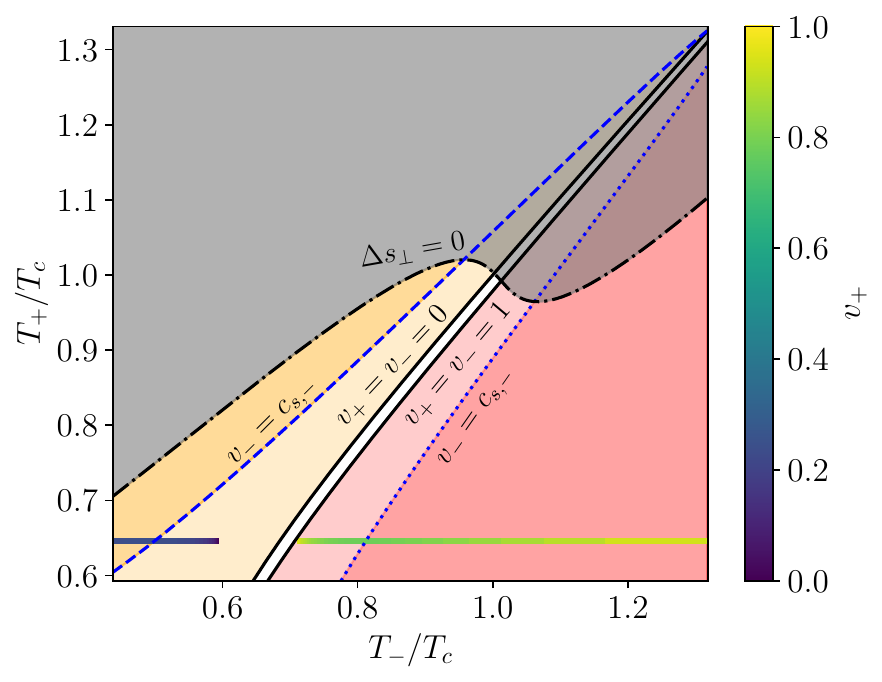}%
    \includegraphics[width=0.5\textwidth]{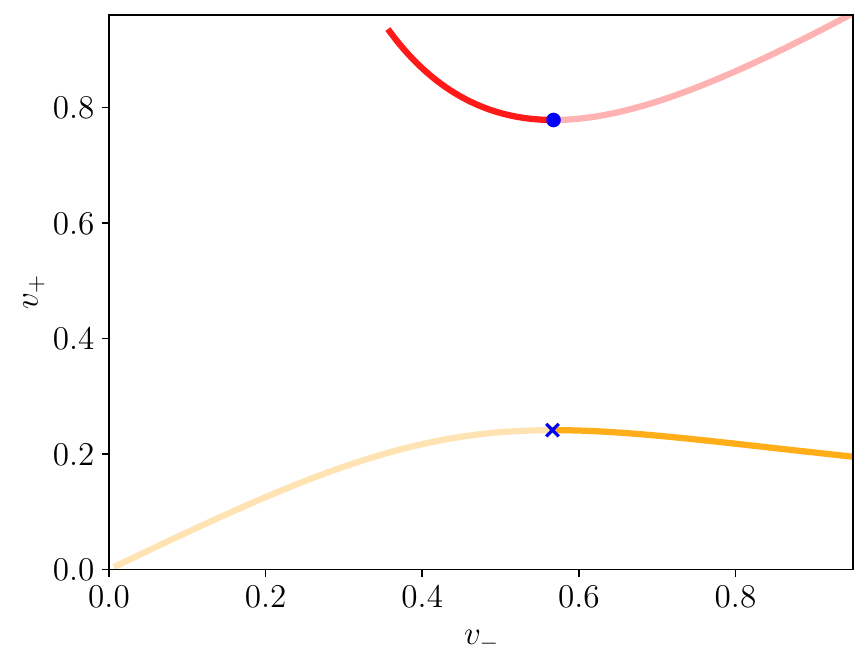}
    \caption{\textbf{Left}: The $T_+$-$T_-$ plane of the EoS~\eqref{eq:EoSf} and plots of solutions of \eqref{eq:vline} and \eqref{eq:jline} for $\mathrm{BP}_1$ (top) and $\mathrm{BP}_2$ (bottom).
    The grey shaded regions are excluded by entropy production. 
    The orange shaded region, lying below the $\Delta s_\perp = 0$ line (black dashed) and above the $v_+=v_-=0$ line (black solid), represents deflagrations.
    The dark orange subregion marks strong deflagrations, and the light orange subregion marks weak deflagrations.
    The region lying below the $\Delta s_\perp = 0$ line and to the right of the $v_+=v_-=1$ line (black solid) marks detonations.
    The dark red subregion shows strong detonations, and the light red subregion shows weak detonations.
    The blue dashed and blue dotted lines are the Jouguet lines.
    The horizontal gradient colour bars represents the varying of $v_+$ and $T_-$ for different $v_-$ at $T_+ = T_n$.
    \textbf{Right}: The two brunches of solutions of $v_+$ as a function of $v_-$ for $T_+ = T_n$. 
    The upper lines are detonations, and the lower lines are deflagrations.
    The blue dot and cross are the Jouguet points.
    Weak deflagrations (light orange part) lie to the left of the blue cross, and strong deflagrations (dark orange part) lie to the right of the blue cross.
    Conversely, weak detonations (light red part) lie to the right of the blue dot, and the strong detonations (dark red part) lie to the right of the blue dot.} 
    \label{fig:xsmTTvv}
\end{figure}

In the following, we perform the same analysis for the EoS ~\eqref{eq:EoSf} specific to the xSM model.
We start from defining the following two important reference temperatures:
\begin{itemize}
    \item $T_{\rm max}$: the maximum temperature where the low-temperature phase still exists.
    \item $T_{\rm min}$: the minimum temperature where the high-temperature phase still exists.
\end{itemize}
These temperatures serve as limits for the EoS~\eqref{eq:EoSf}.
In other words, the EoS of the high-temperature phase is only applicable for $T > T_{\rm min}$, and the EoS of the low-temperature is only valid for $T<T_{\rm max}$.
If the temperature is beyond these ranges, the EoS is not valid anymore. 
In table~\ref{tb:bps}, we choose two benchmark parameters sets for the $Z_2$-symmetric real singlet extension of the SM.  The corresponding phase transition parameters are: the nucleation temperature $T_n$, the transition strength parameter $\alpha_n$, the inverse duration of phase transition $\beta$ and its dimensionless form $\beta/H_*$.
In this table, the corresponding $T_{\rm min}$ and $T_{\rm max}$ are also given for these two benchmark points.
With these model parameters, we calculate \eqref{eq:vline} and \eqref{eq:jline} numerically for the EoS~\eqref{eq:EoSf}.

Figure.~\ref{fig:xsmTTvv} shows the results of our analysis, where the upper and lower plots stand for  $\mathrm{BP}_1$ and  $\mathrm{BP}_2$ respectively.
The left panels present the $T_+$-$T_-$ plane of $\mathrm{BP}_1$ and $\mathrm{BP}_2$, depicting the solutions of \eqref{eq:vline} and \eqref{eq:jline}, and grey shaded regions are excluded by entropy production. 
As in the bag model, there are two classes of solutions for the EoS~\eqref{eq:EoSf}: `\textit{deflagrations}' and `\textit{detonations}'.
Similarly to the bag model, deflagrations occupy the region below $\Delta s_\perp = 0$ line (black dashed) and above the $v_+=v_-=0$ line (black solid), and detonations are located below the $\Delta s_\perp = 0$ line and to the right of the $v_+=v_-=1$ line (black solid).
The blue dashed and blue dotted lines are the Jouguet lines, further dividing the detonations and deflagrations into subclasses.
The horizontal gradient colour bars represent the varying of $v_+$ and $T_-$ for different $v_-$ at $T_+ = T_n$.
In the right panel, we show the two branches of solutions of $v_+$ as a function of $v_-$ for $T_+ = T_n$ which correspond to the gradient colour bars shown in the left panel.
The upper branch represents detonations, and it is related to the gradient colour bar to the right of the $v_+=v_-=1$ line. 
While the lower branch denotes deflagrations, and it is related to the gradient colour bar to the left of the the $v_+=v_-=1$ line. 
The blue dot and cross are the Jouguet points which are the intersections of the gradient colour bars with the Jouguet lines given in the left panel.
Note that the Jouguet lines play a distinctive role in both panels of this figure: at fixed 
$T_+$, $v_+$ reaches its maximum for deflagrations and its minimum for detonations along these Jouguet lines.
For the detonation branch shown in the left panel, $v_-$ cannot be smaller than a specific value.
A smaller $v_-$ would require a larger $T_-$ according to eqs.~\eqref{eq:mcond}.
However, $T_-$ must be smaller than $T_{\rm min}$ and then it sets a lower bound on $v_-$.

\subsection{Hydrodynamic solutions}
\label{sec:hydro-sol}
For both the bag model of EoS and the EoS of the $Z_2$ symmetric real singlet extension, the above analyses show that there are two classes of solutions for eq.~\eqref{eq:mcond}, i.e., deflagrations and detonations.
These two classes of solutions can be further divided  into six different solutions according to the outflow velocity $v_-$, as described below.

For deflagrations, one has $v_-$ > $v_+$ in the wall frame, that indicates that the flow velocity increases across the wall.
According to the outflow velocity $v_-$, deflagrations can be further divided into different processes by comparing with the sound speed just behind the bubble wall $c_{s,-}$.
If $v_- = c_{s,-}$, the process is called \textit{Jouguet}. 
If $v_- < c_{s,-}$, the process is \textit{weak}.
Otherwise, the process is called  \textit{strong}.
In figure~\ref{fig:bag} and the left panels of figure~\ref{fig:xsmTTvv}, Jouguet deflagrations are shown by dashed blue lines, dark orange regions indicate strong deflagrations, and light orange regions represent weak deflagrations.

For detonations, we have $v_-$ < $v_+$ in the wall frame,  indicating a velocity decrease across the wall.
This is contrary compared to the deflagrations.  Analogous to deflagrations, by comparing the outflow velocity $v_-$ with the sound speed just behind the wall $c_{s,-}$, detonations can also be classified into three distinct processes: Jouguet ($v_- = c_{s,-}$), \textit{weak} ($v_- > c_{s,-}$), and \textit{strong} ($v_- < c_{s,-}$).  Jouguet detonations are shown by dotted blue lines, strong detonations are the dark red regions, while weak detonations are depicted by the light red regions in figure~\ref{fig:bag} and left panels of figure~\ref{fig:xsmTTvv}.

To construct a full hydrodynamic solution in the rest frame of the plasma, we have to match the discontinuities at the bubble wall with the possible shock discontinuities and continuous fluid profiles to satisfy the boundary conditions.
According to our previous study~\cite{Wang:2023jto}, the system during the phase transition can be divided into three regions: the high-temperature phase, the low-temperature phase, and the bubble wall.
Hence, to obtain the continuous fluid profiles, we need to derive the fluid equation of motion (EoM) for both phases.
Away from the bubble wall, one can neglect the contribution of the scalar field to the total energy momentum tensor and treat the plasma of both phases as perfect fluid.
Therefore, according to the energy-momentum conservation of the perfect fluid, we get
\begin{equation}
    \partial_\mu T^{\mu\nu} = 0 .
\end{equation}
Assuming that the fluid profiles possess spherical symmetry, 
we have
\begin{equation}
    \begin{split}
        \partial_t[(e + pv^2)\gamma^2] + \partial_r[(e + p)\gamma^2v] &= -\frac{2}{r}[(e + p)\gamma^2v]\,,\\
\partial_t[(e + p)\gamma^2v] + \partial_r[(ev^2 + p)\gamma^2] &= -\frac{2}{r}[(e + p)\gamma^2v^2]\,,
    \end{split}
    \label{eq:feom}
\end{equation}
where $r$ denotes the distance from the bubble center, and $t$ represents the elapsed time since nucleation.
Upon reaching steady state, the bubble walls generate self-similar fluid profiles, which maintain their relative shape while scaling with bubble growth.
At the steady state stage, the solutions only depends on $\xi = r/t$, and we can write $\partial_t = - (\xi/t)\partial_\xi$ and $\partial_r = (1/t)\partial_\xi$.
Therefore, the EoM of the fluid become
\begin{equation}
\begin{split}
2\frac{v}{\xi} &= \gamma^2(1 - v\xi)\left[\frac{\mu^2}{c_s^2(T)} - 1\right]\partial_\xi v\,,\\
\partial_\xi w &= w\left[1 + \frac{1}{c_s^2(T)}\right]\mu\gamma^2\partial_\xi v\,,\\
\partial_\xi T &= T\gamma^2\mu \partial_\xi v\,,
\end{split}\label{eq:fluideqs}
\end{equation}
where
\begin{equation}
\mu(\xi, v) \equiv \frac{\xi - v}{1 - \xi v},
\end{equation}
and $c_s^2(T) \equiv (dp/dT)/(de/dT)$ is the sound speed of the plasma.
For an expanding bubble, the fluid at the center of the bubble and far ahead of the bubble wall remains at rest in the plasma frame.
In practice, different hydrodynamic solutions demand different boundary conditions for the EoM of the fluid.
In the next section, we discuss the characteristics of various hydrodynamic solutions and identify those permissible solutions in the context of first-order phase transitions.

\begin{figure}[t!]
    \centering
    \includegraphics[width=1\textwidth]{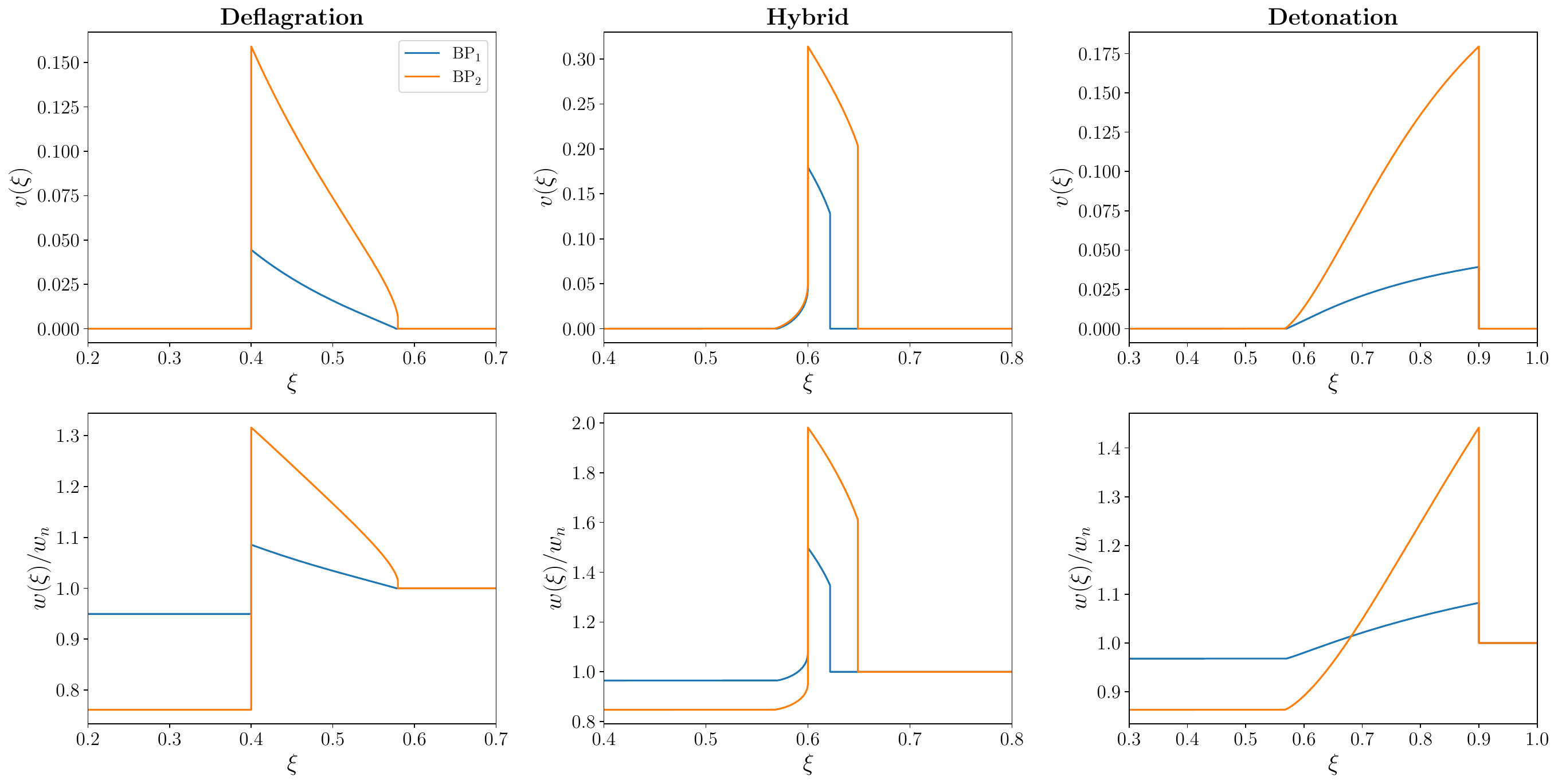}
    \caption{Fluid velocity and enthalpy profiles of $\mathrm{BP}_1$ and $\mathrm{BP}_2$ for deflagration (left), hybrid (middle), and detonation (right).} 
    \label{fig:xsmHd}
\end{figure}

\subsubsection{Deflagrations}

For deflagration solutions, a shock wave can be constructed, yielding a continuous fluid profile between the shock front and the bubble wall. The bubble wall leaves the fluid behind it at rest, while the shock front brings the fluid ahead of it to rest. 
Therefore, the bubble wall velocity of the deflagration solutions $v_w$ coincide with the velocity $v_-$ at which the plasma flows out of the wall, but in the opposite direction. 
Meanwhile, for weak, Jouguet, and strong deflagrations discussed in the last section, the fluid velocity in front of the wall, $v_+$, must be always smaller than the sound speed of the plasma in front of the wall $c_{s,+}$
Additionally, the shock wave can also heat the fluid between the shock front and the bubble wall, resulting in $T_+ > T_n$, where $T_n$ is the nucleation temperature, selected as the reference temperature in this work.
Although there are three different types of deflagrations, not all are physically allowed.
According to the stability argument~\cite{Laine:1993ey}, strong deflagrations are unstable and therefore are not physically viable.
Therefore, deflagration modes studied in this work include only weak and Jouguet deflagrations, which are physically allowed..

To determine the fluid profiles of deflagration solutions, appropriate boundary conditions for the fluid EoM (eq.~\eqref{eq:fluideqs}) must be established.
However, the shock front presents an additional discontinuity for deflagrations, where the fluid velocity should abruptly drop to zero ahead of it. Therefore, the location of the shock front must also be determined.
To resolve this problem, we can employ the energy momentum conservation across the shock front, which gives
\begin{equation}
    v_1 v_2 = \frac{p_1 - p_2}{e_1 - e_2}, \quad \frac{v_1}{v_2} = \frac{e_2 + p_1}{e_1 + p_2}\,.
    \label{eq:sho}
\end{equation}
Here, in the rest frame of the shock front, the subscript $2$ indicates quantities ahead of the shock front, whereas the subscript $1$ denotes quantities behind it, and the shock front velocity is given by $v_{\rm sh} = v_2 = \xi_{\rm sh}$.
Since the shock wave is in the high-temperature phase, the EoS are the same both ahead and behind of the shock front.
Then, a shooting method can be employed to find the fluid profiles of deflagrations.
We apply the shooting method as follows: for a given bubble wall velocity, we have $v_- = v_w$, then we can guess a value for $T_-$, solve eq.~\eqref{eq:mcond} for $T_+$ and $v_+$, and integrate eq.~\eqref{eq:fluideqs} with the initial value $T(\xi = v_w) = T_+$ and $v(\xi = v_w) = \mu(v_w, v_+)$ for $\xi_{\rm sh}$ up to the point where $\mu(\xi_{\rm sh}, v(\xi_{\rm sh}))\xi_{\rm sh} = (p_1(T(\xi_{\rm sh})) - p_2(T_n))/(e_1(T(\xi_{\rm sh})) - e_2(T_n))$ is satisfied.
This procedure is repeated until both equations of~\eqref{eq:sho} are fulfilled.
We present the resulting fluid velocity and enthalpy profiles of the weak deflagration solution in the left panels of figure~\ref{fig:xsmHd} for $\mathrm{BP}_1$ and $\mathrm{BP}_2$ of the $Z_2$ symmetric real scalar singlet extension.

\subsubsection{Detonations}

For detonation solutions, the fluid ahead of the bubble wall is at rest, and a continuous rarefaction wave follows the wall, bringing the fluid back to rest. 
Therefore, the bubble wall velocity of the detonation solutions, $v_w$, corresponds to the velocity $v_+$ at which plasma flows into the wall, but in the opposite direction. 
Since the bubble wall sets the fluid in front of it at rest, we also have $T_+ = T_n$.
Additionally, for weak, Jouguet, and strong detonations, we always have $v_+ > c_{s,+}$.
Strong detonations are not possible solutions for cosmological phase transitions, since the boundary conditions can not be satisfied~\cite{Laine:1993ey,Barni:2024lkj}, which require the fluid to be at rest both at the center of the bubble and far ahead of the bubble wall, and this impossibility is obvious from eq.~\eqref{eq:fluideqs} as explained in the following.

According to the sign convention defined, the fluid velocity $v(\xi)$ (in the plasma frame) within the fluid profiles is always non-negative ($v(\xi) \ge 0$), the boundary conditions enforce the decrease of velocity from $\xi = v_w$ to $\xi = c_s$, where $c_s$ is the sound speed of plasma at the bubble center.
Hence, we have $\partial_\xi v \ge 0$, and these conditions can only be satisfied if 
\begin{equation}
    \left[\frac{\mu^2}{c_s^2(T)} - 1\right] \ge 0\,,
\end{equation}
for the whole interval $[c_s, v_w]$, which indicates $\mu(\xi = v_w, v(\xi = v_w)) = v_- \ge c_{s,-}$ at the wall.
Therefore, we can conclude that strong detonations ($v_- < c_{s,-}$) are impossible for cosmological phase transitions.

To derive the fluid profiles of the other two types of detonations, we need to solve eqs.~\eqref{eq:mcond} to derive $T_-$ and $v_-$, 
with the boundary conditions $v(\xi = v_w) = \mu(v_w, v_-)$ for $\xi = v_w$.
The fluid profiles can be derived by substituting above boundary conditions to eqs.~\eqref{eq:fluideqs}.
We display the resulting fluid velocity and enthalpy profiles of the weak detonation solution in the right panels of figure~\ref{fig:xsmHd}.
Hereafter, detonations modes studied in this work are restricted to be weak or Jouguet detonations.

\subsubsection{Hybrid modes}

Based on the preceding analysis, only weak or Jouguet detonations and deflagrations are viable solutions for expanding bubbles.
Consequently, the bubble wall velocity must be either smaller than $c_{s,-}$ or greater than the Jouguet velocity $v_J$, which is obtained by substituting $v_- = c_{s,-}$ and $T_+ = T_n$ into eq.~\eqref{eq:mcond}, where $v_J > c_{s,-}$.
Thus, for a bubble expands with a wall velocity between $c_{s,-}$ and $v_J$, a pure detonation or a deflagration solution can not be consistently constructed.
For this situation, however, previous studies have shown that a stable solution can be constructed by a Jouguet deflagration with a rarefaction wave behind the bubble wall.
This configuration, a superposition of deflagration and detonation, is usually called hybrid or supersonic deflagration.
In this solution, we have $v_- = c_{s,-}$, and the bubble wall velocity is not directly related to $v_-$ or $v_+$.
Fluid profiles for hybrid solutions can be derived using the presented approach for deflagrations and detonations. 
The middle panels of figure~\ref{fig:xsmHd} illustrate the fluid velocity and enthalpy profiles of the hybrid solution.

\subsection{Bubble wall velocity}

\begin{figure}[t!]
    \centering
    \includegraphics[width=0.5\textwidth]{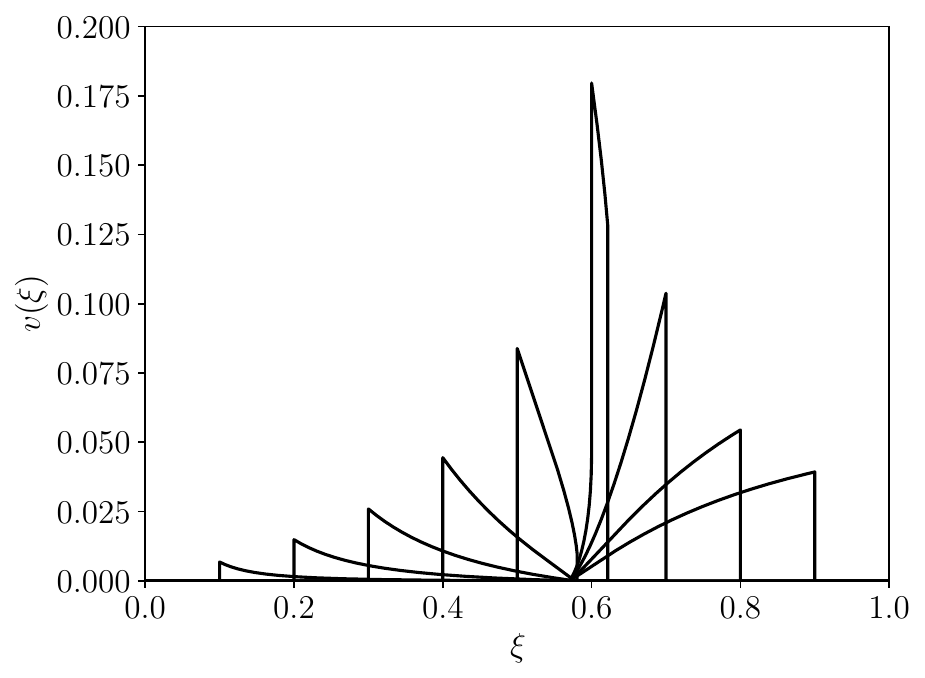}%
    \includegraphics[width=0.5\textwidth]{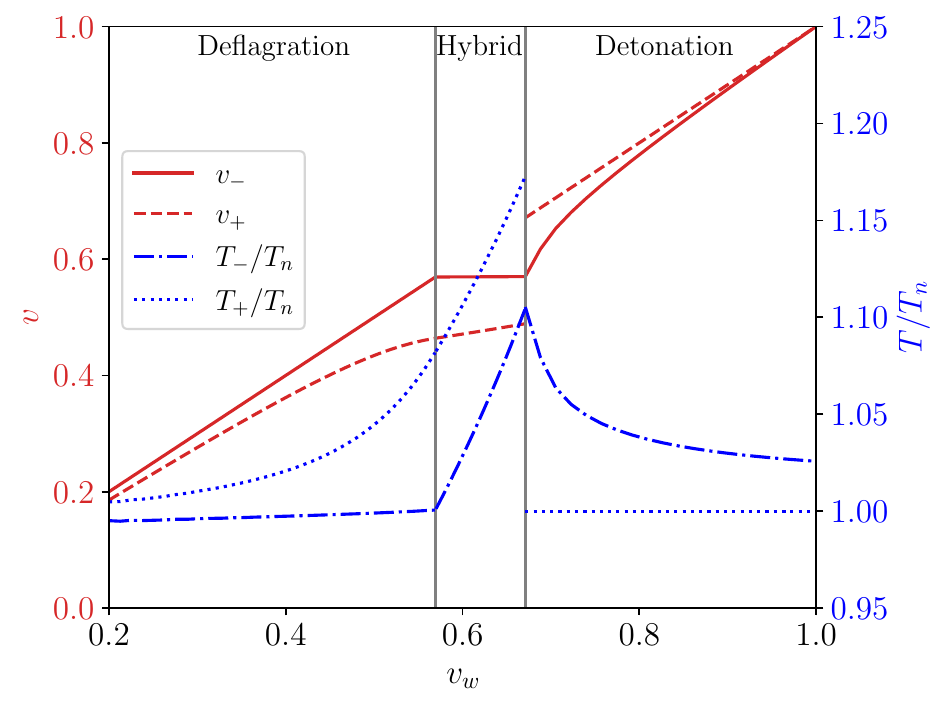}
    \caption{\textbf{Left}: Fluid velocity profiles (in the plasma rest frame) of $\mathrm{BP}_1$ for different bubble wall velocities ($v_w =$ $ 0.1$, $0.2$, $0.3$, $0.4$, $0.5$, $0.6$, $0.7$, $0.8$, $0.9$). \textbf{Right}: Evolution of boundary conditions with the bubble wall velocity for $\mathrm{BP}_1$. Here, $v_\pm$ indicate the fluid velocities in the wall frame.}
    \label{fig:relvw}
\end{figure}

    Based on the above discussion of various hydrodynamic solutions, we can see that the bubble wall velocity is a crucial parameter closely related to the particular hydrodynamical mode of the system.
In this section, we provide a heuristic introduction to the calculation of the bubble wall velocity and demonstrate its correlation with hydrodynamics.

To determine the bubble wall velocity, we must consider the dynamics in the region across the bubble wall, where significant variations in the scalar field make its contribution to the total energy-momentum important.
Therefore, the total energy momentum conservation should be written as 
\begin{equation}
    \partial_\mu T^{\mu\nu} = \partial_\mu (T_{\rm pl}^{\mu\nu} + T_\phi^{\mu\nu} ) = 0\,,
\end{equation}
where 
\begin{equation}
T_\phi^{\mu\nu}  = \partial^\mu\phi_i \partial^\nu\phi_i - \eta^{\mu\nu}\left[\frac{1}{2}\partial_\alpha\phi_i\partial^\alpha\phi_i - V_0(\phi_i)\right]\,,
\end{equation}
is the energy-momentum tensor of $N$ scalar fields $\phi_i$. Here $V_0$ is the zero-temperature part of the effective potential and $\eta^{\mu\nu} \equiv \mathrm{diag}\{+---\}$. 
The energy-momentum tensor of the plasma is
\begin{equation}
    T_{\rm pl}^{\mu\nu} = \sum_i\int\frac{d^3\tilde{\mathbf{p}}}{(2\pi)^3E_{\tilde{p}}}\tilde{p}^\mu \tilde{p}^\nu f_i\, ,
\end{equation}
where $\tilde{p}\equiv(E_{\tilde{p}},\tilde{\mathbf{p}})$ is the 4-momentum.

At the bubble wall, various particle species are out of equilibrium.
Assuming that the deviation from equilibrium is small, the distribution function of particle species $i$ can be expressed as $f_i = f_i^{\rm eq} + \delta f_i$.
Then the energy-momentum tensor of the plasma can be divided into equilibrium and non-equilibrium parts:
\begin{equation}
    T_{\rm pl}^{\mu\nu} = T_{\rm eq}^{\mu\nu} + T_{\rm neq}^{\mu\nu}\, .
\end{equation}
Here
\begin{equation}
    T_{\rm eq}^{\mu\nu} = \sum_i\int\frac{d^3\tilde{\mathbf{p}}}{(2\pi)^3E_{\tilde{p}}}\tilde{p}^\mu \tilde{p}^\nu f_i^{\rm eq}, \quad T_{\rm neq}^{\mu\nu} = \sum_i\int\frac{d^3\tilde{\mathbf{p}}}{(2\pi)^3E_{\tilde{p}}}\tilde{p}^\mu \tilde{p}^\nu \delta f_i,
\end{equation}
and then, the conservation of total energy-momentum tensor can be expressed as
\begin{equation}
    \partial_\mu T^{\mu\nu} = \partial_\mu (T_\phi^{\mu\nu} + T_{\rm eq}^{\mu\nu} + T_{\rm neq}^{\mu\nu})\,.
\end{equation}
Assuming planar symmetry for the wall, in the wall frame, this results in a  system described by the following coupled differential equations:
\begin{align}
    \partial_z^2\phi_i + \frac{\partial V_{\rm eff}}{\partial \phi_i} + \sum_j\frac{\partial(m_j^2)}{\partial\phi_i} \int\frac{d^3\tilde{\mathbf{p}}}{(2\pi)^3E_{\tilde{p}}}\delta f_j &= 0\,,\label{eq:sEOM}\\
    \partial_z\left[w\gamma^2v + F_1\right] &= 0\,,\label{eq:EF1}\\
    \partial_z\left[\frac{1}{2}(\partial_z\phi)^2 - V_{\rm eff} + w\gamma^2v^2 + F_2\right] &= 0\,,\label{eq:EF2}
\end{align}
where $F_1$ and $F_2$ are two quantities derived from $T_{\rm neq}^{\mu\nu}$ (for details see ref.~\cite{Laurent:2022jrs}).
Integration of these equations across the wall should yield eqs.~\eqref{eq:mcond}.
To solve these equations, the corresponding Boltzmann equations for $\delta f_i$ must be solved simultaneously.
For simplicity, we will not delve into solving the Boltzmann equations in this work.
For details on solving  Boltzmann equations, refer to ~\cite{Moore:1995si,Moore:1995ua,Dorsch:2021nje,Dorsch:2021ubz,Dorsch:2023tss,Laurent:2020gpg,Laurent:2022jrs,Wang:2020zlf,Jiang:2022btc,DeCurtis:2022hlx,DeCurtis:2023hil} and our previous work \cite{Wang:2024slx}.

On the other hand, proper boundary conditions for $\phi_i(z)$, $T(z)$, and $v(z)$ are needed for solving these equations.
In general, the boundary conditions for $T(z)$ and $v(z)$ should satisfy
\begin{equation}
    T(-\infty) = T_-,\quad T(+\infty) = T_+,\quad v(-\infty) = v_-, \quad v(+\infty) = v_+,
\end{equation}
and the boundary conditions for $\phi_i(z)$ are
\begin{equation}
    \phi_i(-\infty) = \phi_i^l(T_-), \quad \phi_i(+\infty) = \phi_i^h(T_+),
\end{equation}
where $\phi_i^l(T_-)$ is the VEV of $\phi_i$ in the low-temperature phase at $T_-$, and $\phi_i^h(T_+)$ is the VEV of $\phi_i$ in the high-temperature phase at $T_+$.
These relations indicate that
\begin{equation}
    \frac{\partial V_{\rm eff}}{\partial \phi_i}\Bigg|_{\phi_i = \phi_i^l(T_-)~\mathrm{or}~\phi_i^h(T_+)} = 0\,.
\end{equation}
The $z=\pm\infty$ boundaries should be connected with the fluid profiles of different hydrodynamic solutions.
Different fluid solutions, which are correlated with the bubble wall velocity, yield distinct boundary conditions.
The left panel of figure~\ref{fig:relvw} presents the fluid profiles of $\mathrm{BP}_1$ for varying bubble wall velocities.
As observed, with increasing bubble wall velocity, the velocity profiles transition from deflagration to hybrid and then to detonation.
If the bubble wall velocity $v_w$ is fixed,  the boundary conditions for $T(z)$ and $v(z)$ cannot be chosen freely.
For detonation solutions, we have $v(+\infty) = v_+ = v_w$ and $T(+\infty) = T_n$, whereas deflagration solutions should have  $v(-\infty) = v_- = v_w$, with $T(-\infty) = T_-$ determined consistently.
As for hybrid solutions, we have to utilise the relation $v(-\infty) = v_- = c_{s,-}$ and the relation between the wall and plasma frames to obtain $T(-\infty) = T_-$.
Therefore, the boundary conditions $T_\pm$ and $v_\pm$ vary with the wall velocity. The right panel of figure.~\ref{fig:relvw} illustrates these boundary conditions as functions of the bubble wall velocities for $\mathrm{BP}_1$ of the xSM.

Finally, after establishing the relation between the boundary conditions and the bubble wall velocity and solving the Boltzmann equations, one could employ the shooting method used in ref.~\cite{Wang:2023jto} to solve these equations and obtain the correct bubble wall velocity for a specific particle physics model.
In practice, to simplify the calculation, employing a specific ansatz for $\phi_i$ as done in ref.~\cite{Wang:2020zlf,Jiang:2022btc} is more straightforward.
In this work, we do not calculate the bubble wall velocity for the xSM, instead we treat it as a free parameter.

\subsection{Fluid evolution after collisions}
\label{sec:fl}

To keep track of fluid profiles after the collisions of bubble walls, we employ the philosophy of the so-called hybrid simulation~\cite{Jinno:2020eqg,Jinno:2021ury}.
In this approach, each radial fluid element is assumed to maintain spherical symmetry, and the enthalpy in front of the fluid profiles is shifted to match the value deep inside the bubble after collisions.
This setup incorporates the effect of bubble collisions and  provides the initial conditions for eqs.~\eqref{eq:feom}.
\begin{figure}[t!]
    \centering
    \includegraphics[width=0.5\textwidth]{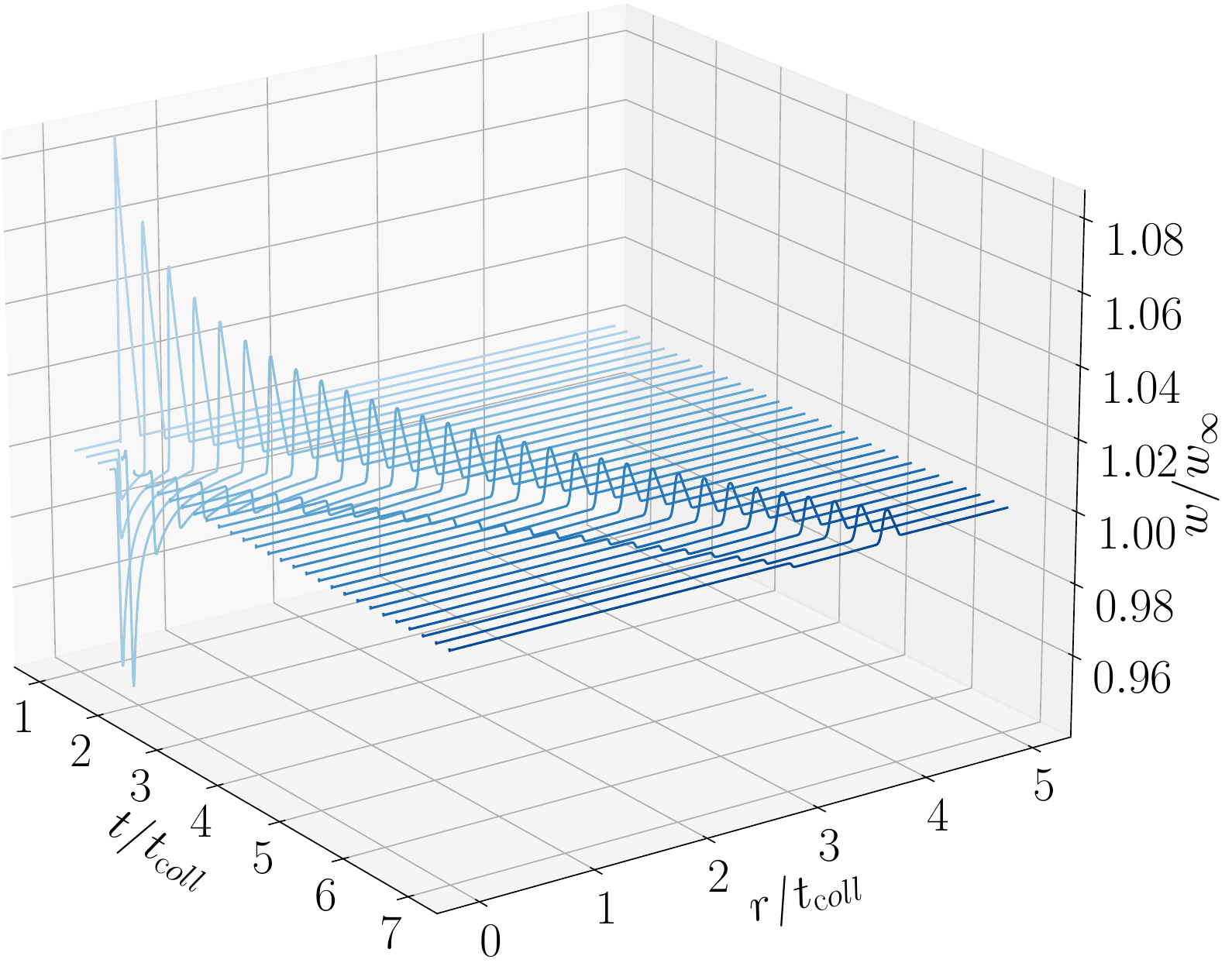}%
     \includegraphics[width=0.5\textwidth]{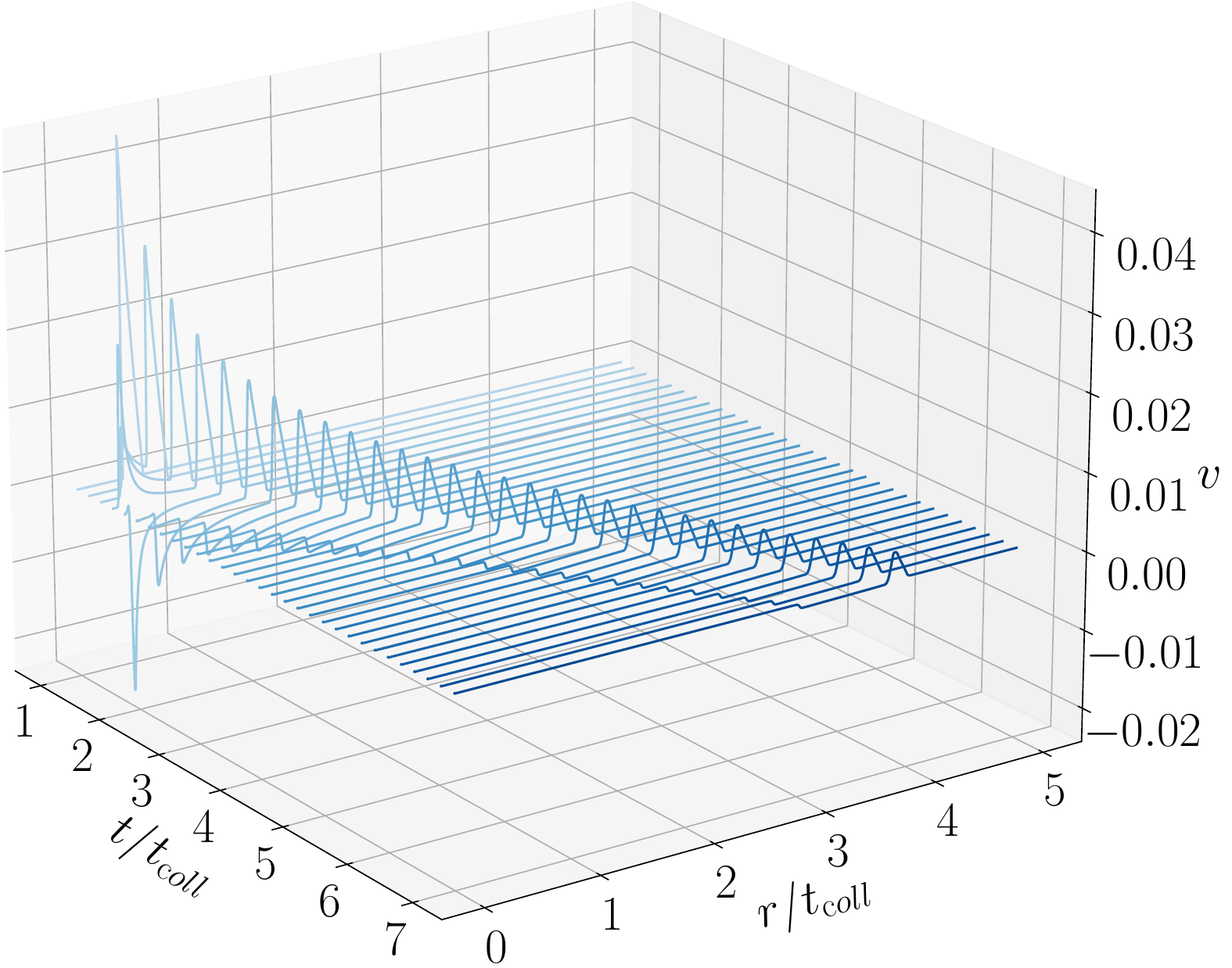}
    \caption{A plot of spherically symmetric fluid profiles after collision for benchmark model $\mathrm{BP}_1$ and $v_w=0.4$. Here $t_{\rm coll}$ denotes the collision time and $w_\infty = w_+(T_n)$.}
    \label{fig:fl_col}
\end{figure}

These spherically symmetric fluid profiles after collisions can be derived by plugging in the relation $w = e+p$ into eq.~\eqref{eq:feom}, yielding equations
\begin{equation}
    \begin{split}
        \partial_t[w\gamma^2 - p] + \partial_r[w\gamma^2v] &= -\frac{2}{r}[w\gamma^2v]\,,\\
\partial_t[w\gamma^2v] + \partial_r[w\gamma^2v^2 + p] &= -\frac{2}{r}[w\gamma^2v^2]\,.
    \end{split}
\end{equation}
These equations can be simplified by introducing conserved variables $Z\equiv w\gamma^2v$ and $E\equiv w\gamma^2 - p$:
\begin{equation}
    \begin{split}
        \partial_t E + \partial_r Z &= -\frac{2}{r} Z\,,\\
        \partial_t Z + \partial_r [Zv + p] &= -\frac{2}{r}Zv\,.
    \end{split}
    \label{eq:EZ}
\end{equation}
The primitive variables are the fluid velocity $v$ and temperature $T$, and they can be obtained from the conserved variables $Z$ and $E$.
In addition, by construction, the following identity holds:
\begin{equation}\label{eq:T&EZ}
    E + p - \frac{1}{2}\left[w + \sqrt{w^2 + 4Z^2}\right] = 0\,.
\end{equation}
Since $p = p(T)$ and $w=w(T)$, we can solve the above equation to derive the temperature $T$ with given $E$ and $Z$.
Then $p$ and $w$ can be directly calculated from the EoS, and the fluid velocity can be derived using the definition of $Z$. 
To better resolve the shock fronts with limited resolution, the Kurganov-Tadmor~\cite{Kurganov:2000ovy} scheme is introduced to evolve these equations.
Figure~\ref{fig:fl_col} illustrates an example of the evolution history of fluid profiles $w$ and $v$ for the benchmark point $\mathrm{BP}_1$ after the collision. 
We observe that wave fronts are well maintained throughout the numerical evolution.

\section{Gravitational waves}
\label{sec:GWs}

Gravitational waves are described by tensor perturbations of the following metric
\begin{equation}
    ds^2 = -dt^2 + a(t)(\delta_{ij} + h_{ij})dx^idx^j\,,
\end{equation}
where $a(t)$ is the scale factor and the irrelevant scalar and vector perturbations are neglected.
If the lifetime of the GW source is much shorter than a Hubble time, we can ignore the expansion of our Universe.  Then the dynamics of the metric perturbation $h_{ij}$ follows the linearised Einstein equations. In momentum space, these equations can be expressed as 
\begin{equation}
    \ddot{h}_{ij} + k^2 h_{ij} = 16\pi G \Lambda_{ij,kl} T^{kl}\,,
\end{equation}
where dots denote a time derivative, $G$ is the Newtonian constant, and $\Lambda_{ij,kl} = P_{ik}P_{jl} - P_{ij}P_{kl}$ with $P_{ij} = \delta_{ij} -\hat{k}_i\hat{k}_j$.  Here $\Lambda_{ij,kl} $ serves as the projection operator employed to extract the transverse-traceless component of the energy-momentum tensor.  From the context the difference between the momentum $k=|\vec{k}|$ and the Lorentz index $k=1,2,3$ is clear. 
These equations have a particular solution for $h_{ij}$, namely
\begin{equation}
    h_{ij}(t, \vec{k}) = 16\pi G\int_0^t dt'\frac{\sin[k(t -t')]}{k}\Lambda_{ij,kl} T^{kl}(t', \vec{k})\,.
    \label{eq:hij}
\end{equation}
When considering a stochastic gravitational wave background, the energy density can be computed by ensemble averaging over several wavelengths,
\begin{equation}
    \rho_{\rm GW} = \frac{\big\langle \dot{h}_{ij}(t, \vec{k}) \dot{h}_{ij}^*(t, \vec{k}) \big\rangle}{32\pi G}\,, 
    \label{eq:rho}
\end{equation}
and the GW spectrum is defined as the ratio between GW energy density per logarithmic interval and the critical density $\rho_{\rm crit}$
\begin{equation}
    \Omega_{\rm GW}^* (t, \vec{k}) \equiv \frac{d \rho_{\rm GW}}{\rho_{\rm crit} d\ln k} \,.
\end{equation}
With eq.~\eqref{eq:hij} and eq.~\eqref{eq:rho}, we can derive 
\begin{equation}
    \Omega_{\rm GW}^* (t, \vec{k})  \propto \big\langle \Lambda_{ij,kl} T^{ij} T^{*kl}\big\rangle\,.
\end{equation}
This suggests that the spectrum of GW energy fraction is generally proportional to the ensemble average of the square of the transverse-traceless projected energy-momentum tensors. In this work, we calculate this spectrum from energy-momentum tensors inspired by the sound waves of the FOPT.

To this end, we assume that the plasma behaves as a perfect fluid during the FOPT, then the corresponding energy momentum-tensor for the GWs is given by 
\begin{equation}
    T^{\mu\nu} = wu^\mu u^\nu + p\eta^{\mu\nu} ,
    \label{eq:swEMT}
\end{equation}
where $u^\mu = \gamma (1, \mathbf{v})$ is the fluid four-velocity, and $\eta^{\mu\nu}\equiv\mathrm{diag}\{-+++\}$ is the Minkowski metric.
Based on this energy-momentum tensor, the GW spectrum can be calculated by the semi-analytical method, e.g. the \textit{Sound shell model}~\cite{Hindmarsh:2016lnk,Hindmarsh:2019phv,Guo:2020grp,Wang:2021dwl,Cai:2023guc,RoperPol:2023dzg,Giombi:2024kju}, or the numerical simulation, e.g,
the \textit{Hybrid simulation}~\cite{Jinno:2020eqg,Jinno:2021ury}, and the \textit{Higgsless simulation}~\cite{Jinno:2022mie,Blasi:2023rqi}.
In this study, to estimate the resulting GW spectra while considering a realistic EoS, we generalize the Hybrid method by making it compatible with the EoS given by eq.~\eqref{eq:EoSf} . Our methodology is detailed as follows.

Adopting the factorisation given in ref.~\cite{Jinno:2020eqg}, the GW spectrum can be expressed as
\begin{equation}
\label{eq:GWsp}
    \Omega_{\rm GW}^*  =  \frac{8\pi G q^3}{4\pi^2\rho_{\rm crit} \mathcal{V}} \int\frac{d\Omega_k}{4\pi}\left[\Lambda_{ij,kl}T_{ij}(q,\vec{k})T^*_{kl}(q,\vec{k})\right]_{q=|\vec{k}|} \simeq \frac{4H\tau_{\rm sw}}{3\pi^2}\frac{H}{\beta}Q'\, .
\end{equation}
Here $Q'$ is the dimensionless growth rate of the GW spectrum defined as
\begin{equation}
\label{eq:qp}
    Q'(q) = \frac{q^3\beta}{w^2\mathcal{V}\mathcal{T}}\int\frac{d\Omega_k}{4\pi}\left[\Lambda_{ij,kl}T_{ij}(q,\vec{k})T^*_{kl}(q,\vec{k})\right]_{q=|\vec{k}|}\,,
\end{equation}
where $w$ denotes the enthalpy of the high-temperature phase, $q$ is the angular frequency, $\vec{k}$ the 3-momentum of the GW waves, $\mathcal{V}$ the simulation volume, and $\mathcal{T}$ the simulation time.
Further, we can estimate the lifetime of the sound waves $\tau_{\rm sw}$ by using the following equation 
\begin{equation}
    H\tau_{\rm sw} = \min\left[1, \frac{HR_*}{\bar{U}_f} \right]\approx \min\left[1, \frac{HR_*}{\sqrt{K}} \right] \,,
\end{equation}
where the mean fluid velocity $\bar{U}_f$ could be approximated by the kinetic fraction $K$ from the self-similar profiles of a single bubble.
Using eq.~\eqref{eq:mbR},  the GW spectrum from the sound waves can be estimated from
\begin{equation}
    \Omega_{\rm GW}^* = \frac{8}{3\pi^{5/3}}\frac{v_w}{\sqrt{K}}\left(\frac{H}{\beta}\right)^2Q'\,.
    \label{eq:GWsw}
\end{equation}

According to the definition of the GW spectrum in eq.~\eqref{eq:GWsp}, 
tracing the evolution of the spatial components of the energy-momentum tensor is crucial in our generalized hybrid method for deriving the GW spectrum produced by sound waves. 
Due to the transverse-traceless projector $\Lambda_{ij,kl}$, only the first term of eq.~\eqref{eq:swEMT} contributes to the GW spectrum.
Therefore, one can estimate $T^{ij}$ as 
\begin{equation}
    T^{ij}(t, \vec{x}) = w(t, \vec{x})\gamma^2(t, \vec{x})v^i(t, \vec{x})v^j(t, \vec{x})\,.
\end{equation}
Ideally, computing this anisotropic energy-momentum tensor requires a full 3-dimensional hydrodynamic simulation. However, in this study, the "hybrid" scheme \cite{Jinno:2020eqg} employed makes it possible to generate a fully anisotropic stress tensor solely based on a spherically symmetric 1-dimensional simulation. This approach requires solving eqs.~\eqref{eq:fluideqs} before bubble collisions and eqs.~\eqref{eq:EZ} after the collisions.
In the following subsection, we will discuss details for the construction of such an anisotropic $T^{ij}$ and the corresponding GW spectrum.

\subsection{Construction of anisotropic $T^{ij}$}

According to Hybrid scheme, the anisotropic and time dependent energy-momentum tensor $T^{ij}(t, \vec{x})$ can be constructed by projecting the results from 1D hydrodynamic simulations with spherical symmetry onto a 3D Cartesian lattice.  
The spherically symmetric fluid profile is obtained by seeking hydrodynamic solutions before and after collisions, as described in Section~\ref{sec:Hydro}. Note that  the realistic EoS (eqs.~\eqref{eq:EoSf}) is considered when solving for these profiles.
To accomplish the 1D to 3D projection, we work within a 3D box with a size that scales with $v_w$: $\mathcal{V} = L^3 = (40 v_w/\beta)^3$, and randomly nucleate bubbles with the nucleation rate given in eq.~\eqref{eq:AppNR} following the strategy described in \cite{Enqvist:1991xw}.  
The entire nucleation process completes at approximately $t\approx 10/\beta$, resulting in the formation of around $2000$ bubbles. 
These bubbles expand and their surrounding fluid profiles during the expansion are assumed to follow the spherically symmetric solutions we have obtained from eqs.~\eqref{eq:fluideqs} and eqs.~\eqref{eq:EZ}.  
After the collision, the spherically symmetric fluid profiles are evolved until $t=6/\beta$. 
Extrapolation is carried out if the profiles beyond this evolution period are needed.

\begin{figure}[t!]
    \centering
    \includegraphics[width=0.33\textwidth]{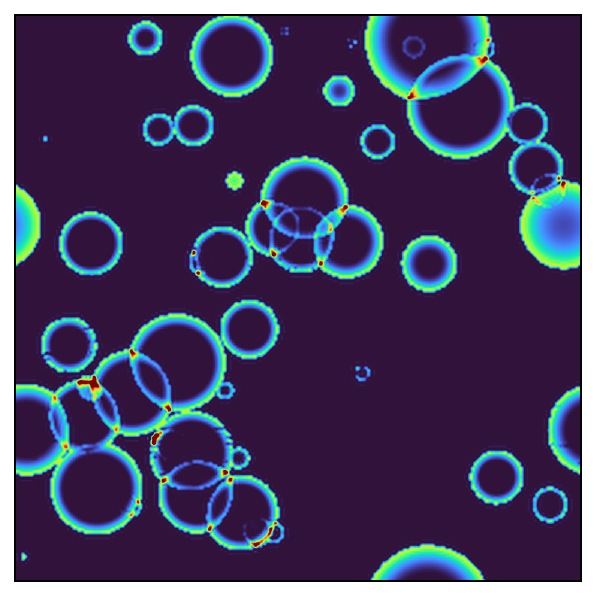}%
    \includegraphics[width=0.33\textwidth]{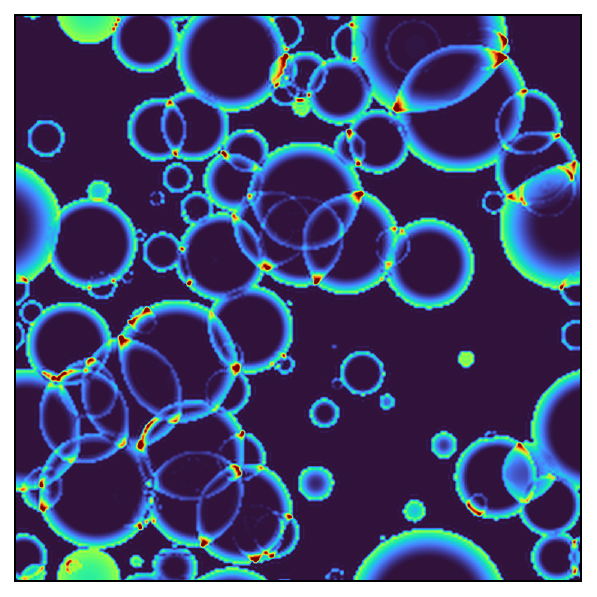}%
    \includegraphics[width=0.33\textwidth]{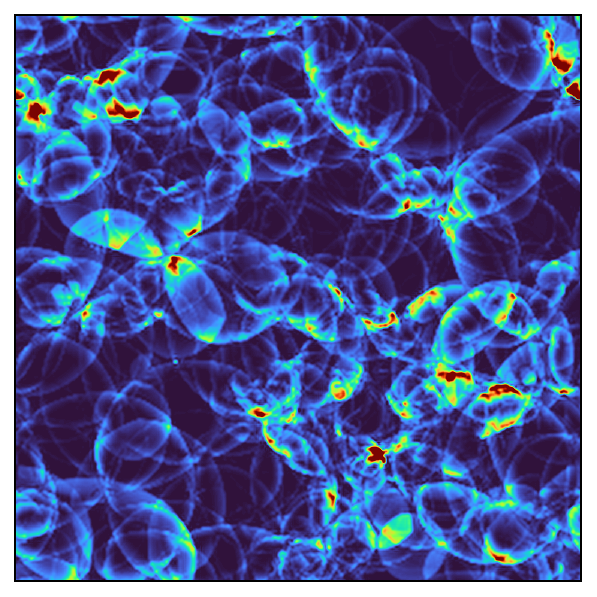}
    \caption{Snapshots of the time evolution of the kinetic energy density $w \gamma^2 v^2$ in our simulations.  These figures correspond to a simulation with $v_w=0.8$ for the benchmark parameter set $\mathrm{BP}_1$ in Table \ref{tb:bps}, with a grid size $N^3=256^3$ and a simulation volume $\mathcal{V} = (40v_w/\beta)^3$. }
    \label{fig:bb}
\end{figure}
In practice, the projection is implemented by superposing enthalpy and velocity profiles of every individual bubbles at each time slice, as follows:
\begin{equation}
    \frac{\Delta w}{ w_0} (t, \vec{x}) \simeq \sum_{i:\rm bubbles} \frac{\Delta w _{\vec{n}_i}}{w_0}(t, |\vec{x} - \vec{n}_i|),\quad \vec{v}(t, \vec{x}) \simeq  \sum_{i:\rm bubbles}  \vec{v}_{\vec{n}_i}(t, |\vec{x} - \vec{n}_i|),
    \label{eq:ghydro}
\end{equation}
where the right hand side of the equations only relies on the spherically symmetric fluid profiles, and the vector $\vec{n}_i$ represents the nucleation position vector for bubble $i$. 
Furthermore, $w_0$ is the enthalpy deep inside the bubble and $\Delta w = w - w_0$.
In fact, eqs.~\eqref{eq:ghydro} are based on the perturbation assumption which might be only solid for phase transitions without strong supercooling.
It is important to decide the direction-dependent collision time $t_{\rm coll}$~\cite{Jinno:2020eqg} for each spherically symmetric radial fluid elements.
When $t < t_{\rm coll}$, the self-similar profiles should be substituted into eq.~\eqref{eq:ghydro}, whereas the profiles given by eqs.~\eqref{eq:EZ} should be used for $t > t_{\rm coll}$.
Here, as an illustration, the result of the projected kinetic energy density $w\gamma^2v^2(t, \vec{x})$ for a detonation mode of $\mathrm{BP}_1$ is presented in Fig. \ref{fig:bb}. 
Note that these spherically symmetric fluid profiles are obtained by solving fluid equations considering a realistic EoS~\eqref{eq:EoSf} (see \ref{sec:fl}). 
Therefore, following the projection, the GW spectrum obtained by this method approximately represents the spectrum predicted by the corresponding particle physics model. 

After obtaining the full anisotropic $T^{ij}(t, \vec{x})$ on the 3D lattice, the dimensionless growth of the GW spectrum $Q'$ can be computed from eq. \eqref{eq:qp}, where the frequency dependence in $T^{ij}(q, \vec{k})$ can be connected to our $T^{ij}(t, \vec{x})$ via Fourier transformation:
\begin{equation}
    T^{ij}(q, \vec{k}) = \int dt e^{iqt} \int d^3 x  e^{-i\vec{k} \cdot \vec{x}} T^{ij}(t, \vec{x}).
\end{equation}
More details about the 1D to 3D projection and computation of the GW spectrum can be found in Appendix B of ref.~\cite{Jinno:2020eqg}. After obtaining $Q'$, we use eq. \eqref{eq:GWsw} to convert it to the GW spectrum created by sound waves.

\subsection{GW spectra}

\begin{figure}[!t]
    \centering
    \includegraphics[width=0.33\textwidth]{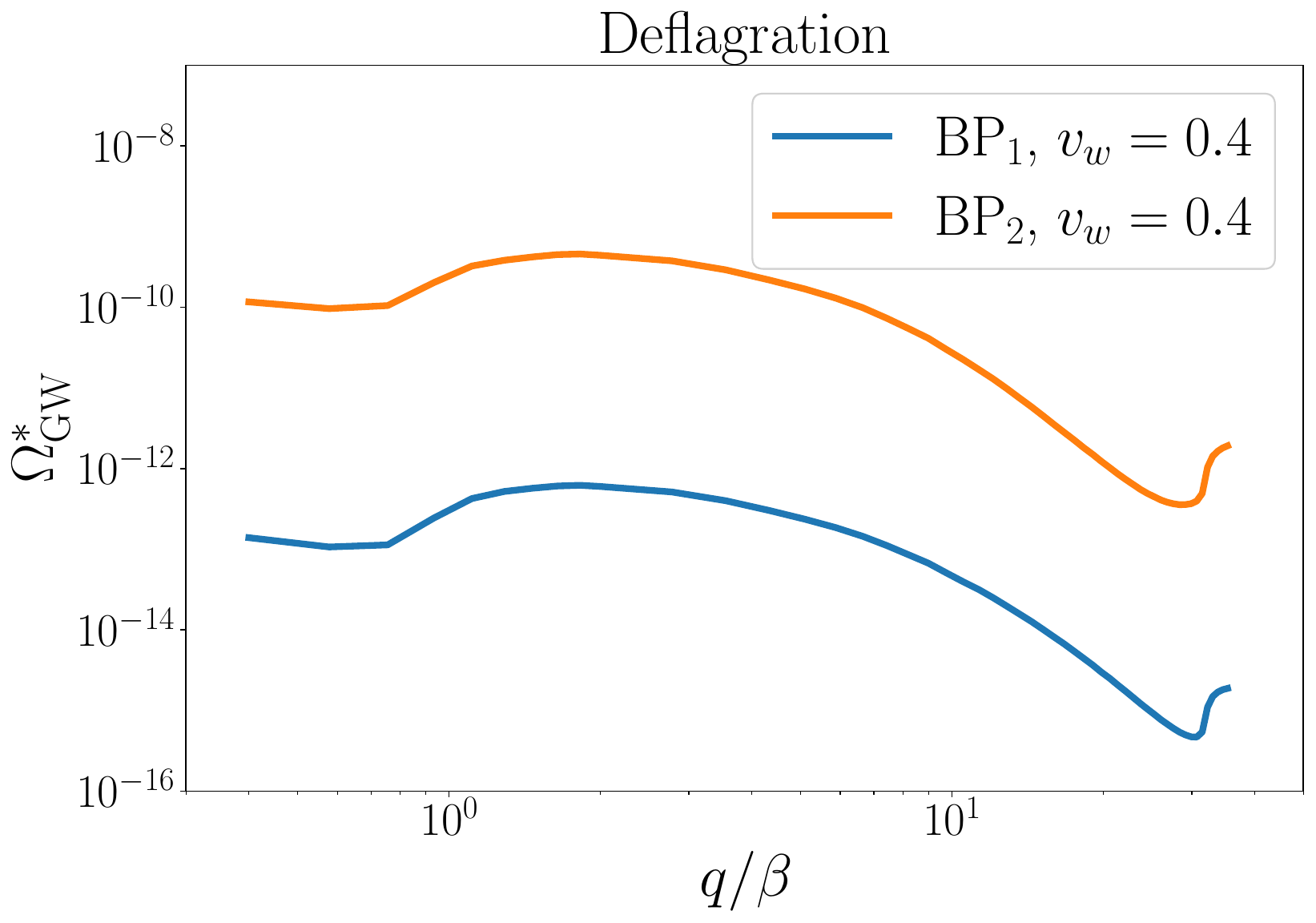}%
    \includegraphics[width=0.33\textwidth]{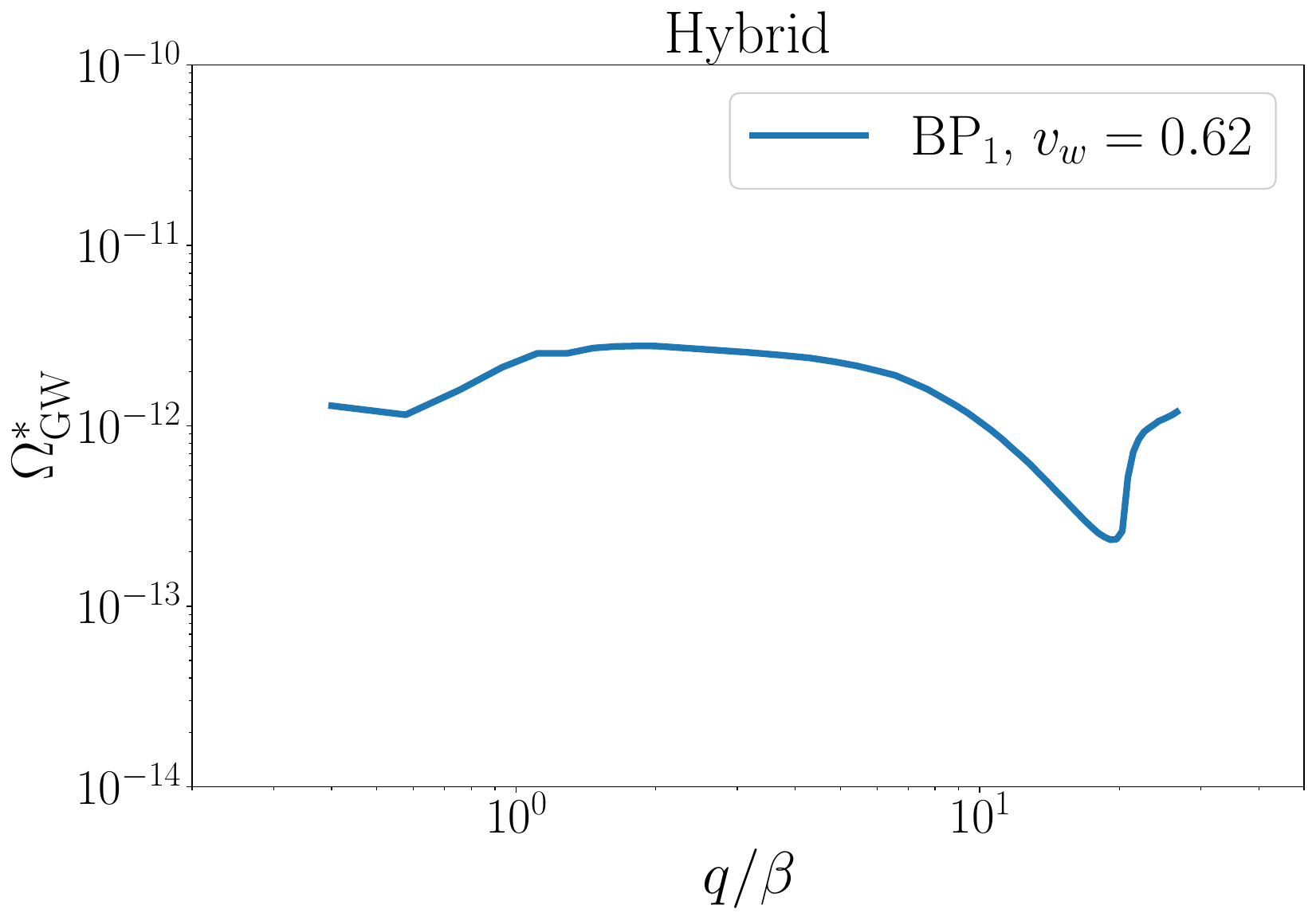}%
    \includegraphics[width=0.33\textwidth]{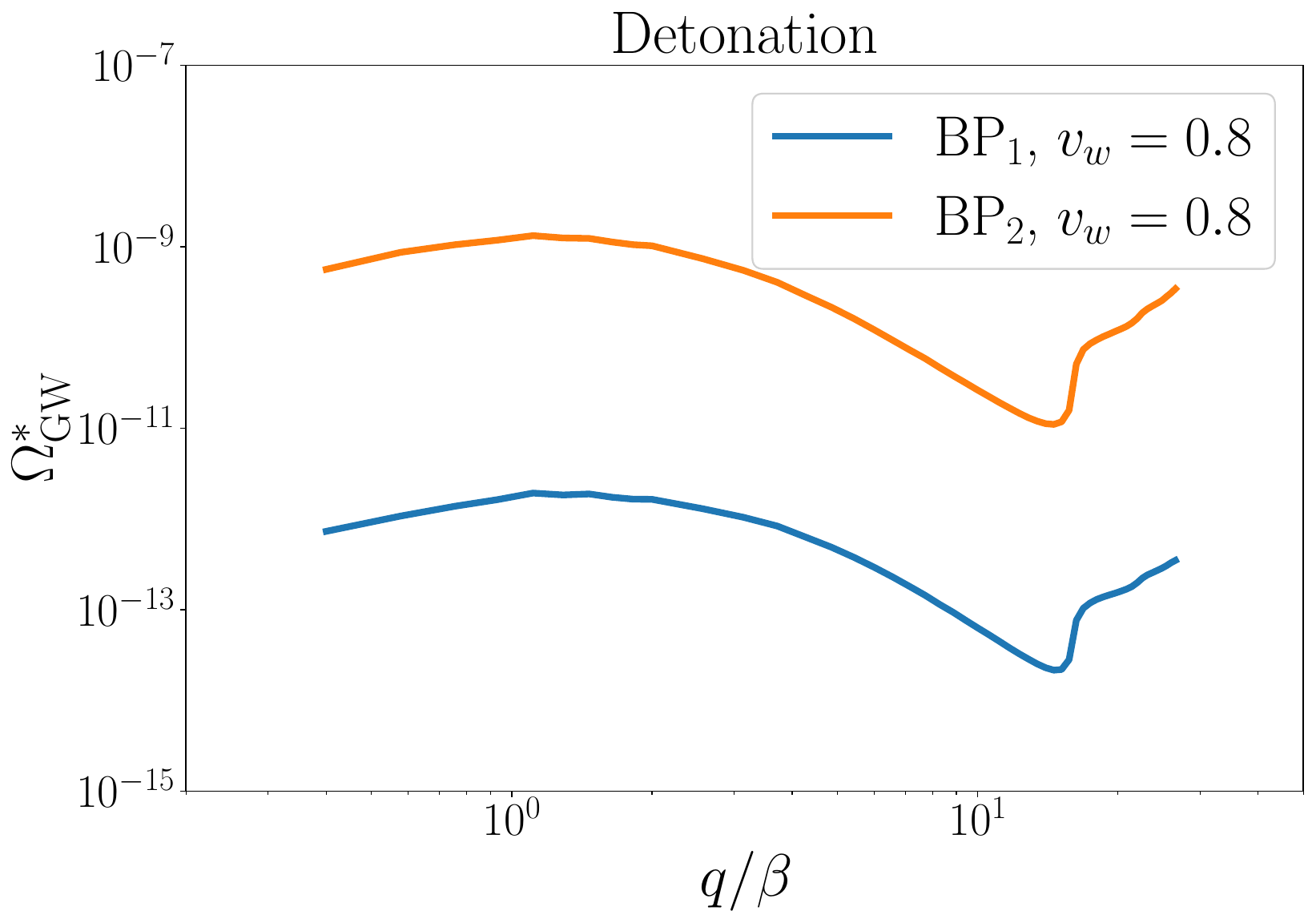}
    \caption{Gravitational wave spectra at production time for the benchmark models in Table \ref{tb:bps} with different wall speeds.}
    \label{fig:om_star}
\end{figure}

We compute the GW spectra $\Omega_{\rm GW}^*$ at the time of production for our benchmark models, $\mathrm{BP}_1$ and $\mathrm{BP_2}$, with three different modes (excluding the hybrid mode for $\mathrm{BP}_2$ as our scheme is unable to obtain reliable results; see section~\ref{sec:concl} for a detail discussion) inspired by various wall speed, $v_w$. 
The results presented in Fig. \ref{fig:om_star} show that while the shapes of $\Omega_{\rm GW}^*$ are similar for the same wall speed, model $\mathrm{BP}_2$, which has a stronger phase transition, produces GWs with larger amplitude. 
It appears that the GW spectrum suddenly increases when $q/\beta$ is larger than around $20$, resulting in rising tails at high frequencies.
Since these high-frequency tails are likely caused by the effect of finite grid size in the simulation, we cut them off when presenting the GW spectrum after cosmological redshift.

\begin{figure}[!t]
    \centering
    \includegraphics[width=0.49\textwidth]{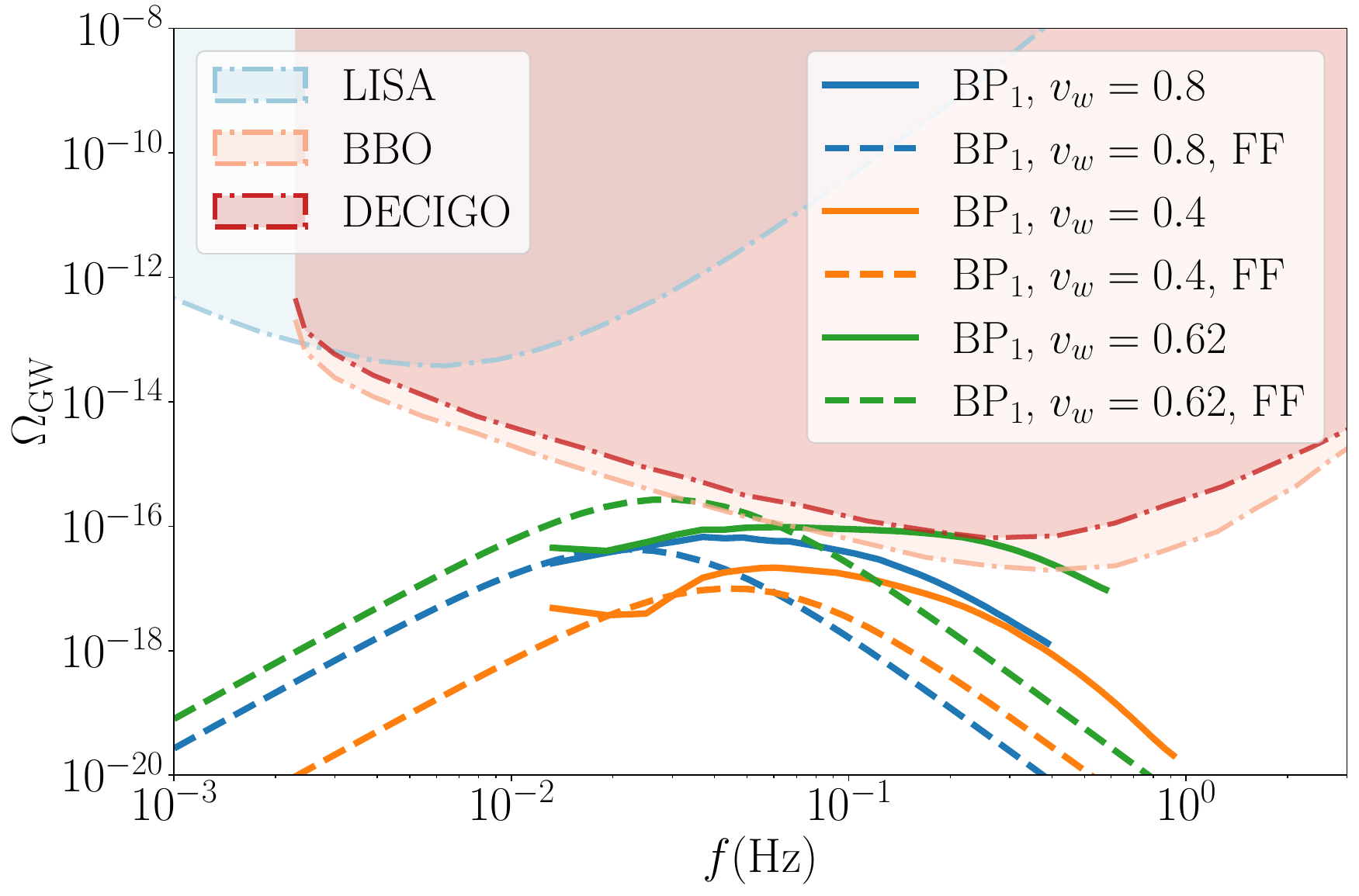}%
    \includegraphics[width=0.49\textwidth]{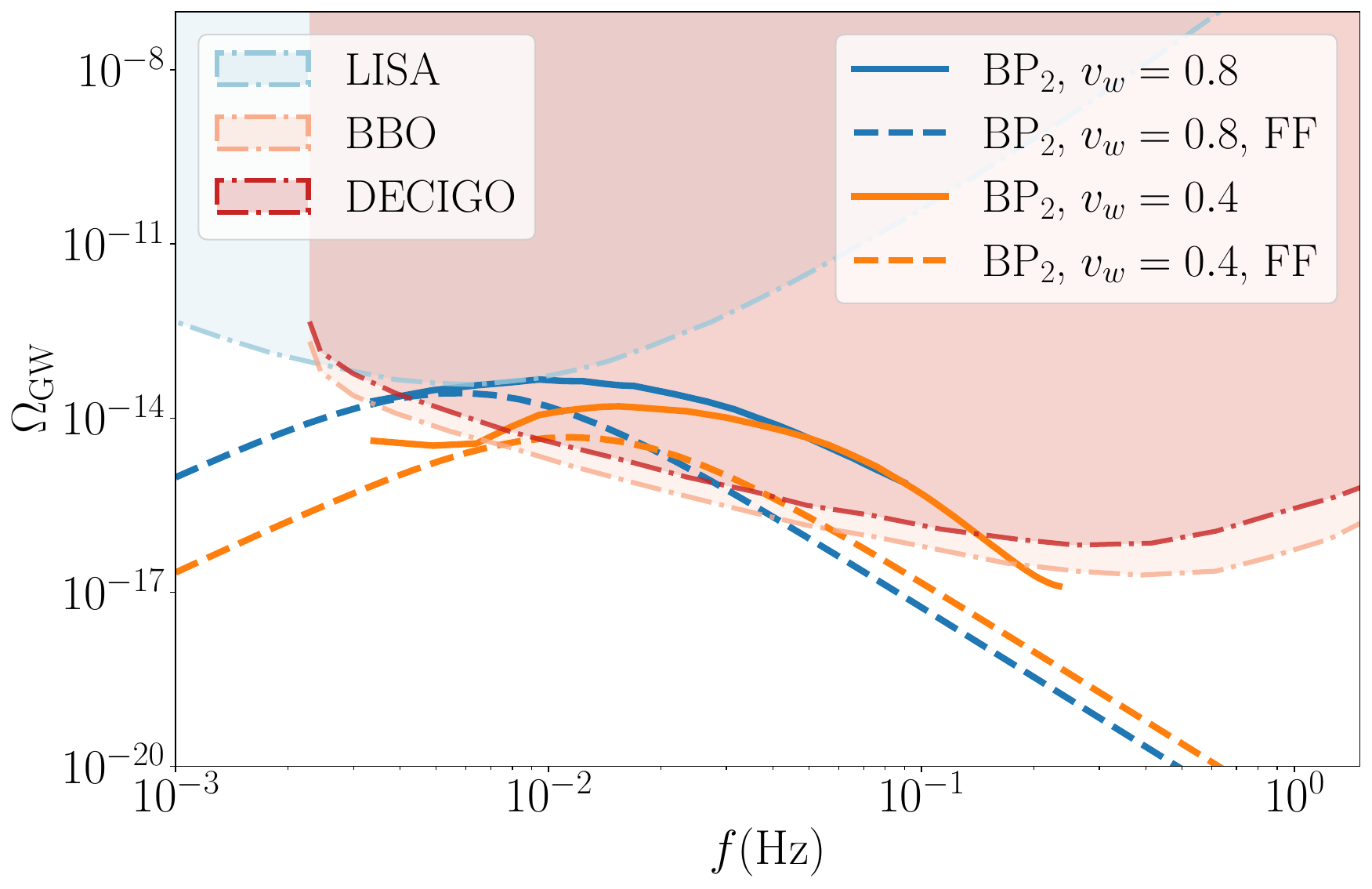}
    \caption{Gravitational wave spectra predicted by the benchmark models in Table \ref{tb:bps} with different wall speeds. Shaded regions represent the power-law-integrated sensitivity~\cite{Thrane:2013oya,Schmitz:2020syl} of various planned GW detectors. Predictions of spectra from the fitting formulas obtained by the scalar+fluid lattice simulation are also shown by dashed curves.}
    \label{fig:GWs}
\end{figure}

Similarly to \cite{Caprini:2019egz}, the GW spectrum today, after the cosmological redshift, is calculated via a rescaling of $\Omega_{\rm GW}^*$ according to:
\begin{equation}
    \Omega_{\rm GW} = F_{\rm GW}\Omega_{\rm GW}^*, 
\end{equation}
where
\begin{equation}
    F_{\rm GW} = (3.57 \pm 0.05) \times 10^{-5}\left(\frac{100}{g_*}\right)^{1/3}\,,
\end{equation}
with redshifted frequencies given by
\begin{equation}
    f = 2.6\times10^{-6}\mathrm{Hz}\frac{q}{\beta}\frac{\beta}{H}\frac{T}{100\mathrm{GeV}}\left(\frac{g_*}{100}\right)^{1/6}.
\end{equation}
We show in Fig.~\ref{fig:GWs} the GW spectra generated by our benchmark models listed in  Table \ref{tb:bps}. As mentioned earlier, the hybrid mode is ignored for model $\mathrm{BP}_2$. We find that GW spectra given by benchmark model $\mathrm{BP}_1$ mainly lie within the sensitivity range of BBO or DECIGO. In constrast, GWs from $\mathrm{BP}_2$ are marginally stronger, typically exhibiting lower frequencies and higher strength, placing them within the LISA sensitivity region.  The higher GW amplitude is a nature consequence of the stronger phase transition strength of the BP2 model. 
We also notice that the spectrum derived by the hybrid solutions tend to possess the largest amplitude and a flatter slope for intermediate frequency band. 

Moreover, based on the phase transition parameters computed from our benchmark models, the GW spectra given by a widely used fitting formula (FF) (see \cite{Espinosa:2010hh} ) are also presented in figure~\ref{fig:GWs} as dashed curves.  
This fitting formula is based on the \textit{Scalar field}+\textit{fluid lattice simulation}~\cite{Hindmarsh:2013xza,Hindmarsh:2015qta,Hindmarsh:2017gnf}, which gives the relation between the GW spectrum and phase transition parameters as:
\begin{equation}
\Omega_{\rm sw}(f) \simeq 3.62\times10^{-6}(H\tau_{\rm sw})(HR_*)\left(\frac{\kappa_v\alpha}{1 + \alpha}\right)^2\left(\frac{100}{g_*}\right)^{1/3}(f/f_{\rm sw})^3\left(\frac{7}{4 + 3(f/f_{\rm sw})^2}\right)^{7/2},\label{eq:swf}
\end{equation}
with the peak frequency
\begin{equation}
f_{\rm sw} \simeq 2.6 \times10^{-5}\text{Hz}\frac{1}{H_*R_*}\left(\frac{T_*}{100 \rm GeV}\right)\left(\frac{g_*}{100}\right)^{1/6} \, .
\end{equation}
Here $\kappa_v = \kappa_v(\alpha,v_w)$ is the efficiency parameter, which can be obtained with the method shown in ref.~\cite{Espinosa:2010hh}. 
As shown in figure~\ref{fig:GWs}, although our model predicts similar amplitude of GWs in the lower frequency range, the FF generally predicts a spectrum with a distinctly different shape across the accessible frequency range compared with the spectrum obtained by using our method.
In general, the FF is based on a broken power-law spectrum, whereas our simulations suggest that the GW spectrum should exhibit a pattern more closely resembling to the double broken power-law.
Furthermore, the GW spectra obtained by our method for both benchmark models in deflagration modes suggest the presence of more complicated behaviour in the low-frequency range.
However, due to our limited computational resources during this project, we are not able to fully explore the low-frequency behaviour with a larger simulation box. 
Recent studies~\cite{RoperPol:2023dzg,Sharma:2023mao} have shown that the GW spectrum generated by sound waves possesses more complicated features at low frequency.
We will leave a detailed study for these IR features of the GW spectra for the future.

\section{Conclusions}
\label{sec:concl}

In this work, we present a detailed analysis of the hydrodynamics and a sophisticated calculation of the gravitational waves generated during the electroweak symmetry breaking with the $Z_2$-symmetric real singlet extension of the standard model.  Specifically, we provide a model-dependent calculation of the GW spectrum based on the framework proposed in ref~\cite{Wang:2024slx}.  By systematically incorporating particle physics model dependencies throughout the calculation, this framework reduces theoretical uncertainties in predicting GWs from phase transitions. This is achieved through deriving an EoS governing the hydrodynamics of the first-order phase transition directly from the effective potential, obtained using conventional 4D perturbative calculations. With this EoS, we are able to perform a more precise hydrodynamic analysis both before and after the bubble collisions.  Finally, we demonstrate that by generalising the original hybrid simulation, the GW spectra for this model are then obtained.

Compared to GW spectra derived from fitting formulas, our method generally produces spectra  with higher peak frequencies and shapes  more closely resembling to a double-broken power law.  For detonations, the magnitude of the spectra from both methods are generally in reasonable agreement.
However, for deflagration or hybrid modes, our method predicts spectra with amplitudes that are 2 to 3 times larger. 

Note that for relatively strong first-order phase transitions, obtaining reliable results for hybrid hydrodynamic solutions after bubble collisions is challenging for our scheme.  Specifically, after a certain period of evolution following the collision, we find that the temperature around $r = 0$ exceeds the valid range for the low-temperature phase of the EoS.  As a result, we were unable to derive the corresponding GW spectrum for the hybrid hydrodynamic mode in $\mathrm{BP}_2$.  We propose three potential explanations for this issue: either the KT scheme fails to provide accurate results for this specific mode, the consequences of collisions are not properly estimated, or this behaviour is physically valid, and more comprehensive calculations are required to confirm this phenomenon.  To draw a definitive conclusion, it will be necessary to explore alternative numerical schemes for solving eqs.~\eqref{eq:EZ} and develop an improved approximation for bubble collisions, which we plan to address in future work.

In summary, by applying the framework from ref.~\cite{Wang:2024slx} to the $Z_2$-symmetric real singlet extension of the SM and conducting a more systematic hydrodynamic analysis, we conclude that our method could quantify GW spectra resulting from this particle physics model more precisely and enable more efficient parameter scans.  Moreover, our method is in principle applicable to other particle physics model.
In the future, this method could be used to investigate model-independent features of GWs generated by sound waves, contributing to a more robust, model-independent description of PTGWs.

\acknowledgments
X.W. would like to thank Henrique Rubira for enlightening discussions.
C.T. is supported by the National Natural Science Foundation of China (Grants No. 12405048) and the Natural Science Foundation of Anhui Province (Grants No. 2308085QA34).
X.W. and C.B. are supported by Australian Research Council grants DP210101636, DP220100643 and LE21010001.

\appendix
\section{Field dependent masses}
\label{App:mass}

The field-dependent mass of various particles in the real singlet extension of SM with $Z_2$ symmetry are given in the following.

\subsection*{Scalars}
The field-dependent mass matrix of scalar particles without thermal corrections is
\begin{equation}
M^2 = \left(\begin{array}{ccc} 
m_{G_\pm}^2 & \boldsymbol{0} &\boldsymbol{0}\\
\boldsymbol{0} & m_{G_0}^2 &\boldsymbol{0}\\
\boldsymbol{0} & \boldsymbol{0} & M_{hs}^2
\end{array}\right).
\end{equation}
where
\begin{equation}
    m_{G_\pm}^2 = m_{G_0}^2 = -\mu_h^2 + \lambda_h\varphi_h^2 + \frac{1}{2}\lambda_{hs} \varphi_s^2
\end{equation}
and 
\begin{equation}
M_{hs}^2 = \left(\begin{array}{cc} 
-\mu_h^2 + 3\lambda_h\varphi_h + \frac{1}{2}\lambda_{hs}\varphi_s^2 & \lambda_{hs}\varphi_h\varphi_s \\
\lambda_{hs}\varphi_h\varphi_s &  -\mu_s^2 + 3\lambda_{s}\varphi_s^2 + \frac{1}{2}\lambda_{hs}\varphi_h^2
\end{array}\right).
\end{equation}
The thermal mass of the scalar particles are
\begin{align}
	\Pi_h &= \frac{T^2}{24}\left[\lambda_h + \lambda_{hs} + \frac{3}{2}(3g_2^2 + g_1^2) + 6y_t^2\right],\\
	\Pi_s &= \frac{T^2}{12}\left[2\lambda_{hs} + 3\lambda_s\right].
\end{align}
Hence we can derive the physical masses for scalar particles by diagonalizing the following mass matrix
\begin{equation}
\overline{M}_s^2 = \left(\begin{array}{cccc} 
m_{G_\pm}^2 + \Pi_h & 0 & 0 & 0\\
0 & m_{G_0}^2 + \Pi_h & 0 & 0\\
0 & 0 & -\mu_h^2 + 3\lambda_h\varphi_h + \frac{1}{2}\lambda_{hs}\varphi_s^2 + \Pi_h & \lambda_{hs}\varphi_h\varphi_s \\
0 & 0 & \lambda_{hs}\varphi_h\varphi_s  & -\mu_s^2 + 3\lambda_{s}\varphi_s^2 + \frac{1}{2}\lambda_{hs}\varphi_h^2 + \Pi_s\\
\end{array}\right).
\end{equation}

\subsection*{Gauge Bosons}
The field-dependent mass matrix of gauge bosons without thermal corrections is
\begin{equation}
M_g^2 = \left(\begin{array}{cccc} 
M_1^2 & 0 & 0 & 0\\
0 & M_1^2 & 0 & 0\\
0 & 0 & M_1^2 & M_{12}^2\\
0 & 0 & M_{12}^2 & M_2^2\\
\end{array}\right)
\end{equation}
where
\begin{equation}
M_1^2 = \frac{g_2^2}{4}\varphi_h^2,\quad M_2^2 = \frac{ g_1^2}{4}\varphi_h^2 ,\quad M_{12}^2 = -\frac{g_1g_2}{4}\varphi_h^2.
\end{equation}
The thermal corrections for the gauge bosons are
\begin{equation}
\Pi_W = \frac{11}{6}g_2^2T^2,\quad\Pi_B = \frac{11}{6}g_1^2T^2\,,
\end{equation}
where $g_2$ and $g_1$ are gauge couplings.
Hence the physical masses of gauge bosons are eigenvalues of the following matrix
\begin{equation}
\overline{M}_g^2 = \left(\begin{array}{cccc}
M_1^2 + \Pi_W & 0 & 0 & 0\\
0 & M_1^2 + \Pi_W & 0 & 0\\
0 & 0 & M_1^2 + \Pi_W & M_{12}^2\\
0 & 0 & M_{12}^2 & M_2^2 + \Pi_B\\
\end{array}\right)\,.
\end{equation}

\subsection*{Top quark}
The field-dependent mass of top quark is
\begin{equation}
m_t^2 = \frac{y_t^2}{2}\varphi_h^2\,,
\end{equation}
where $y_t$ represents the Yukawa coupling of top quark.

\bibliographystyle{JHEP}
\bibliography{ref.bib} 

\end{document}